\documentclass[prx,twocolumn,floatfix,superscriptaddress,longbibliography,notitlepage]{revtex4-1}

\usepackage{amsfonts}
\usepackage[colorlinks=true,linkcolor=blue,citecolor=red,plainpages=false,pdfpagelabels]{hyperref}
\usepackage{amsmath}
\usepackage{amssymb}
\usepackage{graphicx}%
\usepackage[T1]{fontenc}
\usepackage{tcolorbox}
\usepackage{textcomp}
\usepackage[english]{babel}
\usepackage{cleveref}
\usepackage{todonotes}

\definecolor{rred}{RGB}{219, 48, 122}

\usepackage{amssymb}
\usepackage{latexsym}
\usepackage{mathtools}
\providecommand{\U}[1]{\protect\rule{.1in}{.1in}}

\newcommand{\ket}[1]{| #1 \rangle}
\newcommand{\bra}[1]{\langle #1 |}

\def\U{\mathrm{U}}
\def\A{\mathcal{A}}
\def\B{\mathcal{B}}

\def\D{\mathcal{D}}

\def\H{\mathcal{H}}
\def\U{\mathcal{U}}
\def\I{\mathcal{I}}
\def\L{\mathcal{L}}
\def\M{\mathcal{M}}
\def\N{\mathcal{N}}

\def\P{\mathcal{P}}

\def\T{\mathcal{T}}

\newcommand{\tr}{\operatorname{Tr}}

\allowdisplaybreaks

\begin{document}
\title{Optimal tests for continuous-variable quantum teleportation and photodetectors}
\author{Kunal Sharma}
\affiliation{Hearne Institute for Theoretical Physics, Department of Physics and Astronomy, and Center for Computation and Technology,
Louisiana State University, Baton Rouge, Louisiana 70803, USA}
\affiliation{Joint Center for Quantum Information and Computer Science,~University~of~Maryland,~College~Park,~Maryland~20742,~USA}
\author{Barry C. Sanders}
\affiliation{Institute for Quantum Science and Technology, University of Calgary, Alberta T2N 1N4, Canada}
\author{Mark M. Wilde}
\affiliation{Hearne Institute for Theoretical Physics, Department of Physics and Astronomy, and Center for Computation and Technology,
Louisiana State University, Baton Rouge, Louisiana 70803, USA}
\affiliation{Stanford Institute for Theoretical Physics, Stanford University, Stanford, California 94305, USA}
\pacs{}

\begin{abstract}
Quantum teleportation is a primitive in several important applications, including quantum communication, quantum computation, error correction, and quantum networks. In this work, we propose an optimal test for the performance of continuous-variable (CV) quantum teleportation in terms of the energy-constrained channel fidelity between ideal CV teleportation and its experimental implementation. Work prior to ours considered suboptimal tests of the performance of CV teleportation, focusing instead on its performance for particular states, such as ensembles of coherent states, squeezed states, cat states, etc. Here we prove that the optimal state for testing CV teleportation is an entangled superposition of twin Fock states. We establish this result by reducing the problem of estimating the energy-constrained channel fidelity between ideal CV teleportation and its experimental approximation to a quadratic program and solving it. As an additional result, we obtain an analytical solution to the energy-constrained diamond distance between a photodetector and its experimental approximation. These results are relevant for experiments that make use of CV teleportation and photodetectors.
\end{abstract}

\volumeyear{ }
\volumenumber{ }
\issuenumber{ }
\eid{ }
\date{\today}
\startpage{1}
\endpage{10}
\maketitle

\tableofcontents

\section{Introduction}
Quantum teleportation is a fundamental protocol in quantum information theory with no classical analog \cite{bennett1993teleporting}. It allows for the simulation of an ideal quantum channel by making use of entanglement and classical communication.
Other than teleportation of finite-dimensional states \cite{bennett1993teleporting}, quantum states of fields (e.g., optical modes, the vibrational modes of trapped ions, etc.)~can also be teleported using a protocol called \textit{continuous-variable} (CV) quantum teleportation~\cite{BK98}. 

Ideal CV teleportation of an unknown state is only possible in the unrealistic limit of
noiseless homodyne detection and infinite squeezing in the two-mode squeezed vacuum (TMSV) state \cite{BK98}, the latter being the resource state used for the protocol. In such a theoretical setting, CV teleportation simulates an ideal quantum channel. A more practical strategy involves finite squeezing and unideal detection, which simulates an additive-noise channel on input states, instead of simulating an identity channel \cite{BK98}. Due to these limitations, it is natural to ask the following question: how accurately can ideal CV teleportation be simulated by its noisy experimental implementation?

Prior to our work, theoretical and experimental proposals partially answered this question, by estimating the accuracy in teleporting particular quantum states, such as ensembles of coherent states, squeezed states, etc.~\cite{Tan99,braunstein2000criteria,LB2000,braunstein2001quantum,GG01,JBS02,hammerer2005quantum,AC08,lee2011teleportation,CA14,seshadreesan2015non}. Although these states are relevant for several quantum information processing applications, they do not represent the performance of CV teleportation when the goal is to teleport an arbitrary unknown state. Since teleportation can be used as a gadget in several applications for either teleporting an unknown state or applying a gate on an unknown state, it is important to quantify the worst-case performance of CV teleportation.

In this work, we solve this open problem by determining  an optimal test for characterizing the performance of CV teleportation. In particular, by taking the performance metric to be the energy-constrained channel fidelity \cite{Sh16,SWAT} between ideal CV teleportation and its experimental implementation, we determine an optimal input state that can be used to assess the performance of an experimental implementation. Mathematically, this problem is equivalent to calculating the energy-constrained channel fidelity between the identity channel and an additive-noise channel. 
In this work, we develop numerical and analytical techniques to find exact solutions to the optimization involved in calculating the energy-constrained channel fidelity. A consequence of our findings is that there is now an explicit and optimal experimental procedure for characterizing the performance of CV teleportation.
After Eq.~\eqref{eq:opt-state-ex}, we provide details of such an experimental procedure.

The first main contribution of our paper is the reduction of the problem of calculating the energy-constrained channel fidelity between ideal teleportation and its experimental implementation to a quadratic program over an infinite number of variables. We then define a truncated version of this quadratic program and prove that it is a convex optimization problem. We numerically solve it
using a MATLAB package, which employs the interior-point method \cite{pardalos1996interior}. We also provide analytical solutions by invoking the Karush--Kuhn--Tucker (KKT) conditions \cite{karush2014minima,kuhn1951}. We then argue that these solutions to the truncated versions of the quadratic programs also optimize the energy-constrained channel fidelity defined over an infinite-dimensional, separable Hilbert space, i.e., without any truncation at all. 

One of our main findings is that, among all pure bipartite states, an entangled superposition of a finite number of twin Fock states is optimal for distinguishing ideal CV teleportation from its experimental implementation. Our results thus provide an experimental strategy to verify whether an \textit{unconditional} experimental teleportation with high accuracy is possible~\cite{braunstein2000criteria}. Here, by unconditional teleportation, we imply that teleportation of any unknown, energy-constrained state should be feasible. 
We also discuss why previous proposals based on the teleportation of coherent states or other commonly used states are not suitable for quantifying the performance of unconditional teleportation.  

As an additional result, we analytically find the energy-constrained diamond distance between a photodetector and its experimental approximation. We model the noisy version of a photodetector as a pure-loss channel followed by an ideal photodetector. We also find that number-diagonal states optimize the energy-constrained diamond distance and show that entanglement is not required for the optimal distinguishability of the ideal photodetector from its experimental implementation. 


The rest of our paper is organized as follows. We first review some definitions and prior results that will be employed to understand the main results, including quantum states, channels, and distinguishability measures (Section~\ref{app:state-ch-meas}), bosonic Gaussian states and channels (Section~\ref{app:bosonic-st-ch}), and the Karush--Kuhn--Tucker conditions for convex optimization (Section~\ref{app:KKT-cond}).  We then present our results on an optimal test for CV teleportation. 
After that, we present our results on an optimal test for photodetectors. Finally, we summarize our results and state some open problems. 

In the appendices, we provide proofs for the results presented in the main text of our paper.  In Appendix~\ref{sec:CV-teleportation_supp}, we provide proofs related to our results on an optimal test for the performance of continuous-variable (CV) quantum teleportation. Finally, we provide proofs for our results on photodetectors in Appendix~\ref{app:photo-detector}.

\section{Preliminaries}

\label{sec:prelim}

\subsection{Quantum states, channels, and distinguishability measures}

\label{app:state-ch-meas}

In this section, for completeness, we review definitions of quantum states and channels, as well as some distinguishability measures for them. 
Let $\H$ denote an infinite-dimensional, separable Hilbert space. Let $\T(\H)$ denote the set of trace-class operators, i.e., all operators $M$ with finite trace norm:
\begin{equation}
 \left\Vert M \right\Vert_1 \equiv \tr(\sqrt{M^{\dagger}M}) < \infty.   
\end{equation}
 Let $\D(\H)$ denote the set of density operators acting on~$\H$, i.e., those that are positive semi-definite with unit trace. The trace distance between two quantum states $\rho, \sigma \in \D(\H)$ is given by $\left\Vert \rho - \sigma \right\Vert_1$. The fidelity between $\rho$ and $\sigma$ is defined as follows \cite{U76}: 
\begin{equation}
F(\rho,\sigma) \equiv \left\Vert \sqrt{\rho} \sqrt{\sigma} \right\Vert_1^2 .
\label{eq:def-fidelity}
\end{equation}
  The sine distance between $\rho$ and $\sigma$ is defined as \cite{rastegin2002relative,rastegin2003lower,rastegin2006sine,gilchrist2005distance}
\begin{equation}
  C(\rho, \sigma) \equiv \sqrt{1-F(\rho, \sigma)}. 
\end{equation}
The following inequalities relate  the fidelity and the trace distance  \cite{FG98}:
\begin{align}\label{eq:powers}
1- \sqrt{F(\rho, \sigma)} \leq \frac{1}{2} \left\Vert \rho - \sigma \right\Vert_1 \leq \sqrt{1-F(\rho, \sigma)}~. 
\end{align}

Let $H_A$ denote a Hamiltonian  corresponding to the quantum system $A$. Let $\N_{A\to B}$ and $\M_{A \to B}$ be two quantum channels. Let $R$ denote a reference system. Then
the energy-constrained diamond distance between $\N_{A \to B}$ and $\M_{A \to B}$ is defined for $E \in [0,\infty)$ as \cite{S18, W17}
\begin{multline}\label{eq:energy-constrained-diamond-norm}
\left\Vert \N_{A\to B} - \M_{A\to B}\right\Vert_{\diamond E} \equiv \\
 \sup_{\rho_{RA}: \tr(H_A \rho_{A}) \leq E} \left\Vert \N_{A\to B}(\rho_{RA}) - \M_{A\to B}(\rho_{RA}) \right\Vert_1~,  
\end{multline}
where $\rho_{RA} \in \mathcal{D}(\mathcal{H}_R \otimes \mathcal{H}_A)$, and it is implicit that the identity channel $\mathcal{I}_R$ acts on system $R$. Furthermore, the optimization is over every possible reference system~$R$.
The energy-constrained channel fidelity between two quantum channels $\mathcal{N}_{A \to B}$ and $\mathcal{M}_{A \to B}$ for $E\in [0, \infty)$ is defined as \cite{Sh16,SWAT}
\begin{multline}\label{eq:ec-fidelity}
    F_E(\N_{A\to B}, \M_{A \to B}) \equiv\\
      \inf_{\rho_{RA}: \tr(H_A \rho_{A}) \leq E} F(\N_{A\to B}(\rho_{RA}), \M_{A\to B}(\rho_{RA}))~.
\end{multline}
The energy-constrained sine distance between two quantum channels $\N_{A\to B}$ and $\M_{A\to B}$ for $E \in [0,\infty)$ is defined as \cite{Sh16,SWAT}
\begin{multline}\label{eq:ec-sine-distance}
C_E(\N_{A\to B}, \M_{A\to B}) \equiv  \\
  \sup_{\rho_{RA}: \tr(H_A \rho_{A}) \leq E} C(\N_{A\to B} (\rho_{RA}), \M_{A\to B} (\rho_{RA}))~.
\end{multline}
Although each of the above measures is defined with an optimization over mixed states and arbitrary reference systems, it suffices to optimize over pure states such that the reference system $R$ is isomorphic to the channel input system $A$, as a consequence of purification, the Schmidt decomposition, and data processing.

\subsection{Bosonic Gaussian states and channels}

\label{app:bosonic-st-ch}

This section provides a brief review of bosonic Gaussian states and channels. See \cite{H12, AS17} for further details.

Let $\rho \in \mathcal{D}( \H^{\otimes n})$ denote a density operator corresponding to $n$ bosonic modes,  where $\H^{\otimes n}= \otimes_{i=1}^n \H_i$ and $\H_i$ is the Hilbert space corresponding to the $i$th mode. Let $\hat{x}_i$ and $ \hat{p}_i$ denote the respective position- and momentum-quadrature operators of the $i$th mode. Let
\begin{equation}
\hat{r} \equiv (\hat{x}_1, \hat{p}_1, \dots, \hat{x}_n, \hat{p}_n)^T.
\end{equation}
Then the following commutation relation holds:
\begin{align}
[\hat{r}, \hat{r}^T] = i \Omega~, 
\end{align}
where
\begin{equation}
\Omega  \equiv \bigoplus_{i=1}^{n} \Omega_0, \qquad \Omega_0  \equiv \begin{bmatrix}
0 & 1 \\
-1 & 0
\end{bmatrix}    .
\end{equation}
Furthermore, we define the annihilation operator for the $i$th mode as
\begin{equation}
\hat{a}_i \equiv (\hat{x}_i + i \hat{p}_i)/\sqrt{2}.    
\end{equation}

For $r \in \mathbb{R}^{2n}$, any quantum state $\rho \in \D(\H^{\otimes n})$ can be represented as follows:
\begin{align}
\rho = \frac{1}{(2\pi)^n}\int d^{2n} r~\chi_{\rho} (r) D(r)~,
\end{align}
where $D(r) \equiv \exp(i r^T\Omega\hat{r})$ is the unitary displacement operator and $\chi_{\rho}(r) \equiv \tr(D(-r) \rho )$ is the Wigner characteristic function of the state $\rho$. Let
\begin{align}
\mu_{\rho} & \equiv \langle \hat{r} \rangle_{\rho}     ,\\
V_{\rho} & \equiv \langle \{(\hat{r} - \mu_{\rho}), (\hat{r} - \mu_{\rho})^T \}\rangle_{\rho},
\end{align}
denote the mean vector and  covariance matrix of the state $\rho$, respectively. Then $\rho$ is Gaussian if its characteristic function has the following Gaussian form:
\begin{align}
\chi_{\rho}(r) = \exp\!\left(-\frac{1}{4}r^T \Omega^T V_{\rho} \Omega r + i r^T \Omega^T \mu_{\rho}\right)~.
\end{align}

We now review experimentally relevant examples of bosonic states. A coherent state $\ket{\alpha}$ is an eigenvector of the annihilation operator $\hat{a}$ with eigenvalue $\alpha$, i.e., $\hat{a} \ket{\alpha} = \alpha \ket{\alpha}$, which can also be represented as $ \ket{\alpha} = D(\alpha) \ket{0}$, where $D(\alpha) = e^{\alpha \hat{a}^{\dagger} - \alpha^* \hat{a}}$.
A single-mode thermal state with mean photon number $\bar{n} = 1/(e^{\beta \omega} -1)$ has the following representation in the photon number basis:
\begin{equation}\label{eq:thermalstate}
\theta(\bar{n}) \equiv \frac{1}{1+\bar{n}}\sum_{n=0}^{\infty} \left(\frac{\bar{n}}{\bar{n}+1}\right)^n | n\rangle\!\langle n |~.
\end{equation}

In our paper, we employ entangled superpositions of twin Fock states, which we define as 
\begin{align}\label{eq:twin-Fock-state_supp}
\ket{\psi}_{RA} = \sum_{n=0}^{\infty} \sqrt{p_n} \ket{n}_R\ket{n}_A,
\end{align}
where $p_n \in \mathbb{R}^{+}$ and $\sum_{n=0}^{\infty} p_n = 1$. One special example of such a state is the two-mode squeezed vacuum state with parameter $\bar{n}$, which is equivalent to a purification of the thermal state in \eqref{eq:thermalstate} and is defined as
\begin{equation}
\ket{\psi_{\operatorname{TMS}}(\bar{n})} \equiv \frac{1}{\sqrt{\bar{n}+1}}\sum_{n=0}^{\infty}\sqrt{\left(\frac{\bar{n}}{\bar{n}+1}\right)^n} \ket{n}_R\ket{n}_A~.
\end{equation}

Quantum channels that take an arbitrary Gaussian input state to another Gaussian state are called quantum Gaussian channels. Let $\N$ denote a Gaussian channel that takes $n$ modes to $m$ modes. Then $\mathcal{N}$ transforms the Wigner characteristic function  $\chi_{\rho}(r)$ of a state $\rho$ as follows:
\begin{multline}
\label{eq:Wigner-charac-transformation}
\chi_{\rho}(r) \to 
\chi_{\N(\rho)}(r) = \chi_{\rho}\!\left(\Omega^TX^T\Omega r\right) \times \\
\exp\!\left(-\frac{1}{4}r^T \Omega^T Y \Omega r + i r^T \Omega^T d\right)~,
\end{multline}
where $X$ is a real $2m \times 2n$ matrix, $Y$ is a real $2m \times 2m$ positive semi-definite symmetric matrix, and $d\in \mathbb{R}^{2m}$, such that they satisfy the following condition for $\N$ to be a physical channel:
\begin{align}\label{eq:cptp-Gaussian-condition}
Y + i \Omega - i X \Omega X^T  \geq 0~. 
\end{align}
Furthermore, since a Gaussian state $\rho$ can be completely characterized by its mean vector $\mu_{\rho}$ and covariance matrix~$V_{\rho}$, the action of the Gaussian channel $\mathcal{N}$ on the state $\rho$ can be described as follows:
\begin{align}
&\mu_{\rho} \to X \mu_{\rho} + d~, \nonumber \\
& V_{\rho} \to X V_{\rho} X^T + Y~. 
\end{align}

A quantum pure-loss channel is a Gaussian channel that can be characterized by a beamsplitter of transmissivity $\eta\in (0,1)$, coupling the signal input state with the vacuum state, and followed by a partial trace over the environment. In the Heisenberg picture, the beamsplitter transformation is given by the following Bogoliubov transformation:
\begin{align}
\label{eq:beam-splitter-transformation1}
\hat{b} & = \sqrt{\eta} \hat{a} - \sqrt{1-\eta} \hat{e},\\ 
\hat{e}'  & = \sqrt{1-\eta}\hat{a} + \sqrt{\eta} \hat{e},
\label{eq:beam-splitter-transformation2}
\end{align}
where $\hat{a}, \hat{b}$, $\hat{e}$, and $\hat{e}'$  are the annihilation operators representing the sender's input mode, the receiver's output mode, an environmental input mode, and an environmental output mode of the channel, respectively. Let $\L^{\eta}_{A\to B}$ denote a pure-loss channel with transmissivity $\eta$.
Then the action of the pure-loss channel $\L^{\eta}_{A\to B}$ on a state $\rho_A$ is given by 
\begin{align}
    \L_{A\to B}^{\eta}(\rho) \equiv (\tr_{E'} \circ \B^{\eta}_{AE\to BE'})(\rho_A \otimes |0\rangle\!\langle 0|_E),
\end{align} 
where $\mathcal{B}^{\eta}_{AE\to BE'}$ denotes the beamsplitter channel corresponding to \eqref{eq:beam-splitter-transformation1}--\eqref{eq:beam-splitter-transformation2}.
Moreover, the $X$ and $Y$ matrices for $\mathcal{L}^{\eta}$ are given by $X = \sqrt{\eta} I_2$ and $Y = (1-\eta) I_2$, where $I_2$ is a two-dimensional identity matrix.

A quantum-limited amplifier channel with parameter $G \in (1, \infty)$ is characterized by a two-mode squeezer, coupling the signal input with the vacuum state, followed by a partial trace over the environment. We denote a quantum-limited amplifier channel by $\mathcal{A}^G_{A\to B}$. Then the mode transformation corresponding to the two-mode squeezing transformation is given by
\begin{align}
    \hat{b} &= \sqrt{G} \hat{a} + \sqrt{G-1} \hat{e}^{\dagger}~, \label{eq:2-mode-sq-transformation1}\\
    \hat{e}^{\prime} & = \sqrt{G-1} \hat{a}^{\dagger} + \sqrt{G} \hat{e}~, \label{eq:2-mode-sq-transformation2}
\end{align}
where $\hat{a}$, $\hat{b}$, $\hat{e}$, and $\hat{e}^{\prime}$ are the same as defined above for a pure-loss channel.
Then the action of the quantum-limited amplifier channel $\mathcal{A}^{G}_{A\to B}$ on a state $\rho_A$ is given by 
\begin{align}
    \mathcal{A}^{G}_{A\to B}(\rho) \equiv (\tr_{E'} \circ \mathcal{S}^{G}_{AE\to BE'})(\rho_A \otimes |0\rangle\!\langle 0|_E),
\end{align} 
where $\mathcal{S}^{G}_{AE\to BE'}$ denotes the two-mode squeezing channel corresponding to \eqref{eq:2-mode-sq-transformation1}--\eqref{eq:2-mode-sq-transformation2}. Furthermore, the $X$ and $Y$ matrices for $\mathcal{A}^G_{A \to B}$ are given by $X = \sqrt{G}I_2$ and $Y = (G-1)I_2$, where $I_2$ is again a two-dimensional identity matrix.

Let $\T^{\xi}_{A\to B}$ denote an additive-noise quantum Gaussian channel, defined as
\begin{equation}
\label{eq:additive-noise-channel}
\T^{\xi}_{A\to B}(\rho_A) \equiv \int d^2\alpha \, G_{\xi}(\alpha) \, D(\alpha) \rho_A D(-\alpha)~,
\end{equation}
where 
$
    G_{\xi}(\alpha) \equiv (1/\pi\xi) \exp(-\vert \alpha \vert^2/\xi)
$
is a zero-mean, circularly symmetric complex Gaussian probability density function with variance $\xi >0$ and $
D(\alpha) \equiv \exp(\alpha \hat{a}^{\dagger} - \alpha^* \hat{a} ),
$
denotes a displacement operator, with $\alpha \in \mathbb{C}$.
From \cite{caruso2006one,garcia2012majorization}, it follows that an additive-noise channel $\T_{A\to B}^{\xi}$ can be expressed as a concatenation of a pure-loss channel $\L_{A\to B'}^{\eta}$ with transmissivity $\eta$ followed by a quantum-limited amplifier channel $\A_{B' \to B}^{1/\eta}$ with gain parameter $1/\eta$, where $\xi = (1-\eta)/\eta$. This observation is critical to the developments in our paper.

Let $\N$ and $\M$ be quantum channels that take one input mode to $m$ output modes. Then $\N$ and $\M$ are jointly phase covariant \cite{SWAT} if the following holds: 
\begin{multline}
    \mathcal{N}_{A \rightarrow B}\left(e^{i \hat{n} \phi} \rho e^{-i \hat{n} \phi}\right) = \\
    \left(\bigotimes_{i=1}^{m} e^{i \hat{n}_{i}(-1)^{a_{i}} \phi}\right) \mathcal{N}_{A \rightarrow B}(\rho)\left(\bigotimes_{i=1}^{m} e^{-i \hat{n}_{i}(-1)^{a_{i}} \phi}\right), 
\end{multline} 
\begin{multline}
\mathcal{M}_{A \rightarrow B}\left(e^{i \hat{n} \phi} \rho e^{-i \hat{n} \phi}\right) = \\
\left(\bigotimes_{i=1}^{m} e^{i \hat{n}_{i}(-1)^{a_{i}} \phi}\right) \mathcal{M}_{A \rightarrow B}(\rho)\left(\bigotimes_{i=1}^{m} e^{-i \hat{n}_{i}(-1)^{a_{i}} \phi}\right) ,
    \end{multline}
where $a_i \in \{0,1\}$ for $i\in \{1, \dots, m \}$, and $\hat{n}_i$ is the photon number operator for the $i$th mode. 

\subsection{Karush--Kuhn--Tucker (KKT) conditions for convex optimization}

\label{app:KKT-cond}

In this section, we review the Karush--Kuhn--Tucker (KKT) conditions used in solving convex optimization problems with inequality constraints. This is also critical to the analysis in our paper. Let $x\in \mathbb{R}^n$ and let $f: \mathbb{R}^n \to \mathbb{R}$. Consider the following primal optimization problem:
\begin{equation}
\label{eq:kkt-primal}
\begin{aligned}
\min_{x\in \mathbb{R}^n} \quad & f(x) \\
\text{subject to}\quad  & u_i(x) \leq 0,~  \forall i \in \{1,\dots,k\} \\
& v_j(x) = 0, ~\forall j\in\{1,\dots,l\} \, .\\
\end{aligned}
\end{equation}
Let $L(x, a, b)$ denote a Lagrangian with the following form:
\begin{align}\label{eq:lag}
    L(x, a, b) \equiv f(x) + \sum_{i=1}^k a_i u_i(x) + \sum_{j=1}^l b_j v_j(x),
\end{align}
where $a \equiv (a_1,\dots, a_k)$, $b \equiv (b_1, \dots, b_l)$, and $a_i, b_i \in \mathbb{R}$.

Using the Lagrange dual function: 
\begin{align}\label{eq:gab}
    g(a, b) \equiv \min_{x\in \mathbb{R}^n} L(x, a, b)~,
\end{align}
the dual problem corresponding to the optimization in \eqref{eq:kkt-primal} can be defined as follows (see \cite{boyd2004convex} for a review):
\begin{equation}\label{eq:dual-prob}
\begin{aligned}
\max_{a, b}\quad & g(a,b) \\
\text{subject to} \quad & a_i\geq 0, ~\forall i\in \{1,\dots, k\},
\end{aligned}
\end{equation}
where $b\in \mathbb{R}^l$. 

Let $\tilde{x}$ denote a primal feasible point and let $(\tilde{a}, \tilde{b})$ denote a dual feasible point. Then it is easy to show that $f(\tilde{x})\geq g(\tilde{a}, \tilde{b})$, which is the weak duality condition. To see this, consider the following chain of inequalities:
\begin{align}
    g(\tilde{a}, \tilde{b}) & = \min_{x \in \mathbb{R}^n} f(x) + \sum_{i=1}^k \tilde{a}_i u_i(x) + \sum_{j=1}^l \tilde{b}_j v_j(x) \label{eq:min-lag}\\
    & \leq  f(\tilde{x}) + \sum_{i=1}^k \tilde{a}_i u_i(\tilde{x}) + \sum_{j=1}^l \tilde{b}_j v_j(\tilde{x})\\
    & \leq f(\tilde{x})~. \label{eq:weak-duality}
\end{align}
The first equality follows from \eqref{eq:lag} and \eqref{eq:gab}. The first inequality follows due to the minimization over all $x\in \mathbb{R}^n$ in \eqref{eq:min-lag}. Since $\tilde{x}$ is a primal feasible point, it satisfies $v_j(\tilde{x}) = 0$ for all $j$ and $u_i(\tilde{x}) \leq 0$ for all $i$. Moreover, since $\tilde{a}$ is a dual feasible point, $\tilde{a}_i\geq 0$ for all $i$. Collectively, these conditions imply that $\sum_{i=1}^k \tilde{a}_i u_i(\tilde{x}) \leq 0$ and $ \sum_{j=1}^l \tilde{b}_j v_j(\tilde{x})=0$, which leads to the last inequality. 

The duality gap $f(\tilde{x}) - g(\tilde{a}, \tilde{b})$ provides a way to bound how suboptimal primal and dual feasible points are. Let $f^*$ denote the primal optimal value and $g*$ the dual optimal value. Then the following inequalities hold:
\begin{align}
    f(\tilde{x}) - f^* \leq f(\tilde{x}) - g^* \leq f(\tilde{x}) - g(\tilde{a}, \tilde{b}),
\end{align}
which follow from the weak duality condition in \eqref{eq:weak-duality} and from the definitions of the primal and dual optimal values. Thus we get
\begin{align}\label{eq:zero-dg}
    f^* \in [f(\tilde{x}), g(\tilde{a}, \tilde{b})]  \quad \text{and} \quad g^* \in [f(\tilde{x}), g(\tilde{a}, \tilde{b})]~,
\end{align}
which implies that the optimality  of $f(\tilde{x})$ and $g(\tilde{a},\tilde{b})$ depends on the duality gap. In other words, if the duality gap is zero, $\tilde{x}$ is a primal optimal point and $(\tilde{a}, \tilde{b})$ is a dual optimal point. 

We now describe the KKT conditions for the aforementioned optimization problem, which are necessary conditions, in the sense that if a primal optimal point $x^*$ and a dual optimal point $(a^*,b^*)$ with zero duality gap exist, they  satisfy the KKT conditions. We later will argue when the KKT conditions are also sufficient for the optimality of a solution. The KKT conditions are given by
\begin{equation}
\begin{aligned}
\text{Stationarity condition} \quad &  \partial_xL(x, a, b)\vert_{x^*}=0\\
\text{Complementary slackness} \quad & a^*_i u_i(x^*) =0, \forall  i \in\{1, \dots, k\} \\
\text{Primal feasibility} \quad & u_i(x^*) \leq 0,~\forall i\in \{1, \dots, k\} \\
\quad & v_j(x^*) = 0, \forall j\in \{1, \dots, l\}\\
\text{Dual feasibility} \quad & a^*_i\geq 0, \forall i \in \{1, \dots, k\}
\end{aligned}
\end{equation}\label{eq:kkt-conditions}

We provide a brief proof justifying why primal and dual optimal points with zero duality gap satisfy the KKT conditions.
First note that if $x^*$ is a primal optimal solution, it satisfies the primal feasibility conditions, as a consequence of \eqref{eq:kkt-primal}. Similarly, if $(a^*,b^*)$ is a dual optimal solution, as a consequence of \eqref{eq:dual-prob}, it satisfies the dual feasibility condition. Moreover, the zero duality gap implies that the inequalities in \eqref{eq:min-lag}--\eqref{eq:weak-duality} should be saturated. Therefore, the primal optimal point $x^*$ minimizes $L(x, a^*, b^*)$, which implies that the partial derivative of $L(x, a^*, b^*)$ at $x=x^*$ is equal to zero. In other words, the stationarity condition is satisfied. Finally, the zero duality gap further implies that
\begin{align}\label{eq:cslackness}
    \sum_i a^*_i u_i(x^*) &= 0.
\end{align}
Since $a^*_iu_i(x^*) \leq 0$ for all $i \in \{1, \dots, k\}$, from \eqref{eq:cslackness}, we get that $a^*_i u_i(x^*)=0$ for all $i$. Thus, complementary slackness is satisfied. This completes the proof. 

In our paper, we solve an optimization problem in which $f(x)$ is a convex quadratic function in $x$, and $u_i(x)$ and $v_j(x)$ are linear in $x$. Thus, the Lagrangian in \eqref{eq:lag} is a convex function in $x$. We now argue that for such optimization problems, the KKT conditions are both necessary and sufficient. Let us assume that $\bar{x}$ and $(\bar{a}, \bar{b})$ satisfy the KKT conditions. Then from the stationarity condition in \eqref{eq:kkt-conditions}, we get $\left. \partial_x L(x, \bar{a}, \bar{b})\right \vert_{\bar{x}}=0$. Since $L(x, \bar{a},\bar{b})$ is convex in $x$, we get that
\begin{equation}
 \min_x L(x, \bar{a}, \bar{b}) = L(\bar{x}, \bar{a}, \bar{b}).
\end{equation}
Therefore, from \eqref{eq:gab}, it follows that 
\begin{align}
    g(\bar{a}, \bar{b}) &= f(\bar{x}) + \sum_{i=1}^k \bar{a}_i u_i(\bar{x}) + \sum_{j=1}^l \bar{b}_j v_j(\bar{x})\\
    & = f(\bar{x}),\label{eq:zdp}
\end{align}
where the last equality follows from the KKT conditions, i.e., $\bar{a}_iu_i(\bar{x})=0$ for all $i$ and $v_j(\bar{x})=0$ for all $j$. Thus, the zero duality gap in \eqref{eq:zdp} implies that $\bar{x}$ and $(\bar{a}, \bar{b})$ are respectively primal and dual optimal solutions, as argued in \eqref{eq:zero-dg}. 

\section{Quantifying the performance of continuous-variable quantum teleportation}

Note that ideal CV teleportation induces an identity channel $\I_{A \to B}$ on input states. On the other hand, an experimental implementation of CV teleportation realizes an additive-noise channel $\T^{\xi}_{A\to B}$ with the noise parameter~$\xi$, as defined in \eqref{eq:additive-noise-channel}, which quantifies unideal squeezing and unideal detection in the teleportation protocol~\cite{BK98}.

\subsection{Reduction of energy-constrained channel fidelity to a quadratic program}

In order to quantify the accuracy in implementing CV teleportation, we evaluate the following energy-constrained channel fidelity between $\I_{A \to B}$ and $\T^{\xi}_{A \to B}$:
\begin{multline}
\label{eq:tele-fid1}
    F_E(\I_{A\to B}, \T^{\xi}_{A \to B}) =  \\
      \inf_{\rho_{RA}: \tr(H_A \rho_{A}) \leq E} F(\I_{A\to B}(\rho_{RA}), \T^{\xi}_{A\to B}(\rho_{RA}))~,
\end{multline}
where $E\in [0, \infty)$ denotes the energy constraint, $H_A$ denotes a Hamiltonian corresponding to the quantum system $A$, $\rho_{RA}\in\D(\H_{RA})$ denotes an arbitrary state, and it is implicit that the identity channel $\mathcal{I}_R$ acts on system~$R$. Here, $F(\rho, \sigma)$ is the fidelity as defined in \eqref{eq:def-fidelity}. 

Since the identity channel (ideal teleportation) and an additive-noise channel are jointly phase covariant (recall the definition from  Section~\ref{app:bosonic-st-ch}), it suffices to restrict the optimization in the energy-constrained channel fidelity in \eqref{eq:tele-fid1} over pure states having the following form \cite{SWAT}:
\begin{equation}\label{eq:tps}
\ket{\psi}_{RA} = \sum_{n=0}^{\infty} \sqrt{p_n} \ket{n}_R\ket{n}_A,
\end{equation}
for some $p_n \in \mathbb{R}^{+}$ such that $\sum_{n=0}^{\infty} p_n n \leq E$ and $\sum_{n=0}^{\infty} p_n = 1$. We call the state $\ket{\psi}_{RA}$ in \eqref{eq:tps} an entangled superposition of twin Fock states.

To see the claim above, consider that, as discussed previously, the optimization in \eqref{eq:tele-fid1} can be conducted over pure states satisfying the energy constraint, as follows: 
\begin{align}\label{eq:fid-cv-tel-abstract2-supp}
F_E(\I_{A\to B}, \T^{\xi}_{A\to B}) = \inf_{\phi_{RA}: \tr(\hat{n}_A \phi_{A})\leq E} F(\phi_{RA}, \T^{\xi}_A(\phi_{RA})),
\end{align}
where $\phi_{RA}$ is a pure state. We now argue that the optimization in \eqref{eq:fid-cv-tel-abstract2-supp} can be further restricted to pure states that are entangled superpositions of twin Fock states, satisfying the energy constraint. Let $\phi_{RA}$ be an arbitrary state satisfying the energy constraint, and set $\phi_A = \tr_R(\phi_{RA})$. Consider the following phase averaging of $\phi_A$:
\begin{align}
    \psi_{A} &\equiv \frac{1}{2 \pi} \int_{0}^{2 \pi} d \theta \, e^{i \hat{n} \theta} \phi_{A} e^{-i \hat{n} \theta}\\
    &=\sum_{n=0}^{\infty}\vert n\rangle\!\langle n\vert\phi_{A}\vert n\rangle\!\langle n\vert\\
    & = \sum_{n=0}^{\infty} p_n | n\rangle\!\langle n |_A,\label{eq:psi_A}
\end{align}
where $p_n = \langle n \vert \phi_A \vert n \rangle$. 

Then from isometric invariance and monotonicity of fidelity, and from the joint phase covariance of $\mathcal{I}_{A\to B}$ and $\mathcal{T}^{\xi}_{A\to B}$, it follows that (see Proposition~54 of \cite{SWAT}, as well as \cite{leditzky2018approaches})
\begin{multline}\label{eq:iso-mono-fid}
    F(\psi_{RA}, (\mathcal{I}_R \otimes \T^{\xi}_{A\to B}) (\psi_{RA})) \\
    \leq F(\phi_{RA}, (\mathcal{I}_R\otimes\T^{\xi}_{A\to B})(\phi_{RA})),
\end{multline}
where $\psi_{RA}= |\psi\rangle \! \langle \psi |_{RA}$ is a purification of $\psi_A$ in \eqref{eq:psi_A}.

Since the phase averaging operation does not change the mean photon number, we get that \cite[Proposition~54]{SWAT}
\begin{equation}
    \tr(\hat{n} \psi_A) = \tr(\hat{n} \phi_A) .
\end{equation}
Thus, combining \eqref{eq:fid-cv-tel-abstract2-supp} and \eqref{eq:iso-mono-fid} further reduces the optimization in \eqref{eq:fid-cv-tel-abstract2-supp} as follows:
\begin{multline}\label{eq:fid-cv-tel-abstract3-supp}
    F_E(\I_{A\to B}, \T^{\xi}_{A\to B}) \\
    = \inf_{\psi_{RA}: \tr(\hat{n}_A \psi_{A})\leq E} F(\psi_{RA}, \T^{\xi}_{A\to B}(\psi_{RA})),
\end{multline}
where $\ket{\psi}_{RA}$ is given by \eqref{eq:tps} with 
$\sum_{n=0}^{\infty} p_n n \leq E $.

We now show that the optimization in \eqref{eq:tele-fid1} can be formulated as a quadratic program (see \cite{nocedal2006numerical}  for a review of quadratic programs).
Recall that the adjoint of a quantum-limited amplifier channel $\A^{1/\eta}$ is related to a pure-loss channel $\L^{\eta}$ in the following sense \cite{ivan2011operator}:
$
(\A^{1/\eta})^{\dagger} = \eta \L^{\eta},
$
which leads to
\begin{equation}
    F(\psi_{RA}, \T^{\xi}_{A\to B}(\psi_{RA})) = \eta \tr((\L^{1-\eta}(\psi_A))^2),
    \label{eq:key-eq-fidelity-rewrite}
\end{equation}
where 
\begin{equation}
\label{eq:opt-rho-warbit-lambda_supp}
\psi_A  = \sum_{n=0}^{\infty} p_n | n\rangle\!\langle n |_A    .
\end{equation}
To see the claim in \eqref{eq:key-eq-fidelity-rewrite}, 
recall that the adjoint of a quantum-limited amplifier channel $\A^{1/\eta}$ is related to a pure-loss channel $\L^{\eta}$ as follows \cite{ivan2011operator}:
\begin{align}\label{eq:qamp-adjoint_supp}
(\A^{1/\eta})^{\dagger} = \eta \L^{\eta}. 
\end{align}
Then we get
\begin{align}
    & F(\psi_{RA}, \T^{\xi}_{A\to B}(\psi_{RA})) \notag \\
    &= \tr(\psi_{RA} \T^{\xi}_{A\to B}(\psi_{RA}))\\
    & = \tr(\psi_{RA} (\A^{1/\eta}\circ \L^{\eta})(\psi_{RA}))\\
    & = \eta \tr((\L^{\eta}(\psi_{RA}))^2)\\
    & = \eta \tr((\L^{1-\eta}(\psi_A))^2). \label{eq:fid-equiv-purity_supp}
\end{align}
Finally, from \eqref{eq:fid-equiv-purity_supp}, we get
\begin{align}
& F(\psi_{RA}, \T^{\xi}_{A\to B}(\psi_{RA})) \notag \\
& =  \eta \tr((\L^{1-\eta}(\psi_A))^2)\\
& = \eta \sum_{n, m =0}^{\infty}p_n p_m \sum_{k=0}^{\min\{n, m\}} \binom{n}{k} \binom{m}{k} (1-\eta)^{2k} \eta^{n+m-2k}\\
& =\sum_{n,m=0}^{\infty} p_n p_m \sum_{k=0}^{\min\{n,m\}} \binom{n}{k}\binom{m}{k} \frac{\xi^{2k}}{(1+\xi)^{n+m+1}}, \label{eq:quad-prog-cv-telport_supp}
\end{align}
where we used 
\begin{align}
    \L^{1-\eta}(\psi_A) &= \sum_{n=0}^{\infty} p_n\sum_{k=0}^n \binom{n}{k} (1-\eta)^{k} \eta^{n-k} | k \rangle \! \langle k |_A,\\
\xi &= \frac{1-\eta}{\eta}~.
\end{align}

Let $p = (p_0, p_1, \dots)$. Then from \eqref{eq:quad-prog-cv-telport_supp}, the desired optimization problem in \eqref{eq:fid-cv-tel-abstract3-supp} is equivalent to the following quadratic program in $p$ with inequality constraints:
\begin{equation}
\label{eq:quad_prog_supp}
 F_E(\I_A, \T^{\xi}_A) =
 \left\{ \begin{array}{l l}
    \inf_p &   f(p) \\[6pt]
\text{subject to} & \sum_{n=0}^{\infty} n p_n \leq E \\[3pt]
 & p_n \geq 0, \forall n \in \mathbb{Z}^{\geq 0}  \\[3pt]
 & \sum_{n=0}^{\infty} p_n = 1
\end{array}\right. ,
\end{equation}
where 
\begin{multline}
     f(p) \equiv \\
     \sum_{n,m=0}^{\infty} p_n p_m \sum_{k=0}^{\min\{n,m\}} \binom{n}{k}\binom{m}{k} \frac{\xi^{2k}}{(1+\xi)^{n+m+1}} \, .
     \label{eq:quad-prog-cv-telport}
\end{multline} 
Henceforth, we denote $F(\psi_{RA}, \T^{\xi}_{A\to B}(\psi_{RA}))$ as $f(p)$. 



In general, solutions to quadratic programs can be obtained numerically by using a MATLAB package that employs the interior-point method \cite{pardalos1996interior}. Moreover, analytical solutions can be calculated by invoking the Karush--Kuhn--Tucker (KKT) conditions \cite{karush2014minima,kuhn1951}. However, these methods are suitable for solving optimization problems over a finite number of variables. Therefore, we first define a truncated version of the energy-constrained channel fidelity between two quantum channels, which is equivalent to a quadratic program over a finite number of variables for the task of distinguishing the identity channel from an additive-noise channel. 

Let $M$ denote the truncation parameter, and let $\mathcal{H}_M$ denote an $(M+1)$-dimensional Fock space $\{\ket{0}, \ket{1},\ldots, \ket{M}\}$. Let $\omega_A \in \mathcal{D}(\mathcal{H}_M)$. We define the energy-constrained channel fidelity between $\mathcal{I}_{A\to B}$ and $\mathcal{T}^{\xi}_{A\to B}$ on a truncated Hilbert space as 
\begin{multline}\label{eq:fid-truncated}
    F_{E,M}(\mathcal{I}_{A\to B}, \mathcal{T}^{\xi}_{A\to B})  
  \equiv \\\inf_{\substack{\omega_{RA} \in \mathcal{D}(\mathcal{H}_R \otimes \mathcal{H}_M):\\
  \tr(\hat{n}\omega_A)\leq E}} F(\mathcal{I}(\omega_{RA}), \mathcal{T}^{\xi}(\omega_{RA})),
\end{multline}
where $\omega_{RA}$ is an extension of $\omega_A$. Similar to the previous case, it suffices to optimize over pure bipartite states of systems $R$ and $A$, with system $R$ isomorphic to system~$A$, so that the dimension of $R$ can be set to $M+1$.
We redefine the probability vector as $p=(p_0, \ldots, p_M)$. Then from arguments similar to those used in deriving \eqref{eq:quad_prog_supp}, we find that
\begin{equation}
\label{eq:quad_prog_trunc}
 F_{E,M}(\I_{A \to B}, \T^{\xi}_{A \to B}) = \left\{ \begin{array}{l l}
    \inf_p &  f(p) \\[6pt]
\text{subject to} & \sum_{n=0}^{M} n p_n \leq E, \\[3pt]
& p_n \geq 0,\, \forall n,  \\[3pt]
& \sum_{n=0}^{M} p_n = 1.
\end{array}\right.
\end{equation}

We now provide the argument of \cite{Ludovico2020}, that the objective function in \eqref{eq:quad_prog_trunc} is convex in $p$. Let $A(\xi)$ denote the Hessian corresponding to $f(p)$. By inspecting \eqref{eq:quad-prog-cv-telport}, its matrix elements are given by
\begin{align}
    [A(\xi)]_{n,m} & = \frac{\partial^2}{\partial p_n \partial p_m} f(p) \\
    & = 2 \sum_{k=0}^{\min\{n,m\}} \binom{n}{k}\binom{m}{k} \frac{\xi^{2k}}{(1+\xi)^{n+m+1}}.
\end{align}
We note that $A(\xi)$ can be expressed as follows: 
\begin{equation}
  A(\xi)
  = \sum_{k=0}^M \frac{2\xi^{2k}}{1+\xi} | \Upsilon\rangle\!\langle\Upsilon | \geq 0~,\label{eq:hessian-positivity},
\end{equation}
where
\begin{equation}
\ket{\Upsilon} = \sum_{n=0}^M \binom{n}{k} \frac{1}{(1+\xi)^n}\ket{n},    
\end{equation}
which implies that for an arbitrary value of $M$, the Hessian $A(\xi)$ is positive semi-definite. As a consequence, that the objective function $f(p)$ is convex in $p$ (see \eqref{eq:hessian-1}--\eqref{eq:hessian-4} in Appendix~\ref{app:solve-quad-prog} for more details). 

\begin{figure}
		\includegraphics[
		width=0.9\columnwidth
		]{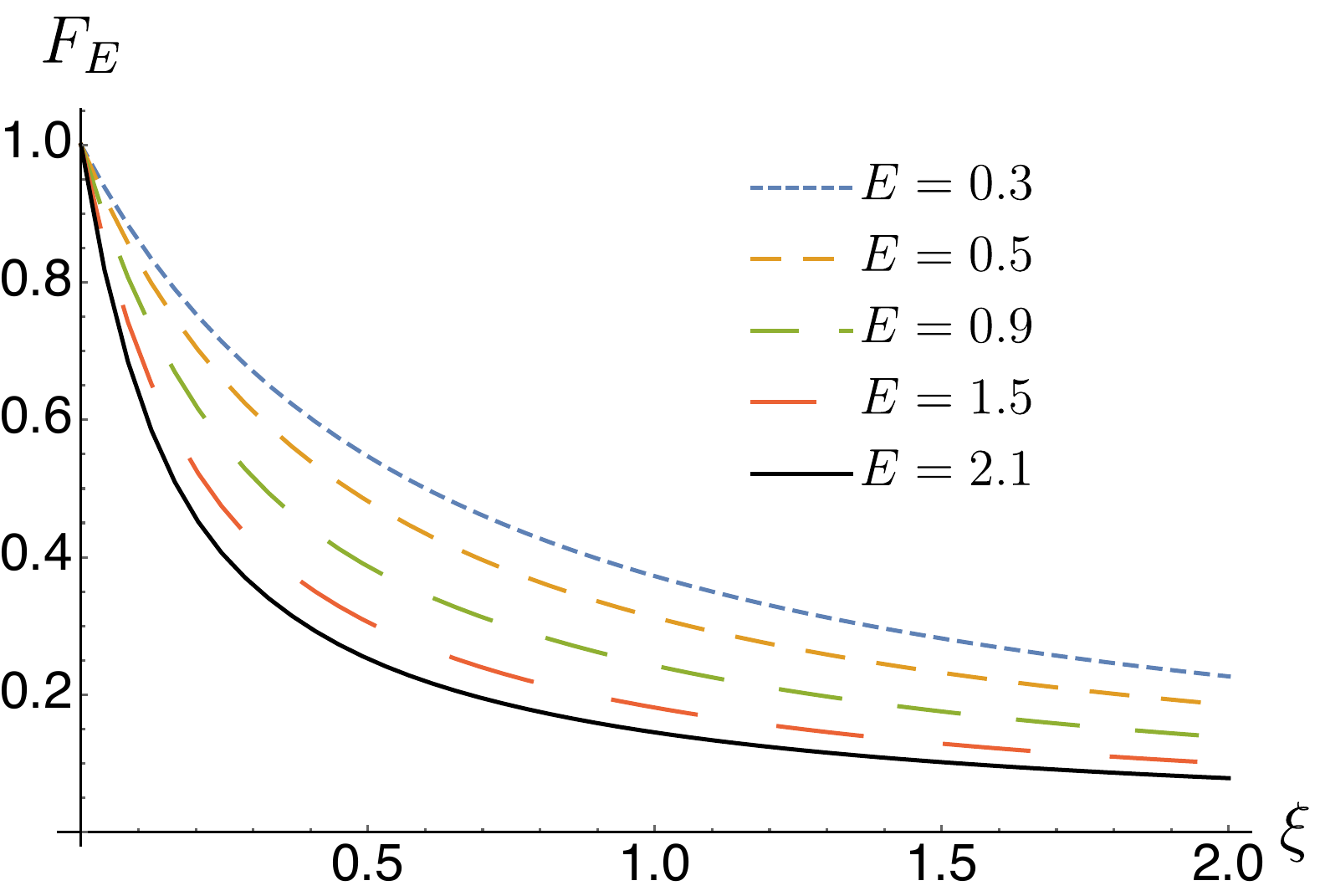}
	\caption{Plot of the energy-constrained channel fidelity $F_E(\I_{A \to B}, \T^{\xi}_{A \to B})$ between ideal teleportation and its experimental implementation versus the noise parameter $\xi$. Here, we denote $F_E(\I_{A \to B}, \T^{\xi}_{A \to B})$ by $F_E$ for simplicity. In the figure, we select certain values of the mean-photon energy $E$ of the channel input, with the choices indicated in the figure legend. For a fixed value of $E$, we solve the optimization program in \eqref{eq:quad_prog_trunc} for $M=50$. The figure indicates that for a fixed value of the noise parameter $\xi$, the simulation of CV teleportation is more accurate as the energy constraint on the input states decreases.}
	\label{fig:fid-teleportation}
\end{figure}

Since the inequality constraints in \eqref{eq:quad_prog_supp} are linear in~$p$, the convexity of $f(p)$ ensures that an optimal point obtained from either numerical or analytical methods is the global optimal point. In Figure~\ref{fig:fid-teleportation}, we plot solutions of the quadratic program in \eqref{eq:quad_prog_trunc} for different values of the energy-constraint parameter $E$, with the choices indicated next to the figure. As shown in Figure~\ref{fig:fid-teleportation}, for a fixed value of the noise parameter $\xi$, the accuracy in implementing CV teleportation decreases as the energy constraint on the input states increases. 

Using the KKT conditions, we can obtain analytical solutions to the optimization problem in \eqref{eq:quad_prog_supp}. For arbitrary values of $E\geq 0 $ and $\xi \geq 0$, the analytical form of the solution can be obtained by first solving the problem numerically for high values of the truncation parameter~$M$, which provides information about the non-zero elements of the optimal probability vector $p^*$. Using that information, the KKT conditions can then be solved analytically.

\subsection{Examples of evaluating the energy-constrained channel fidelity}

We now discuss some examples (please see Appendix~\ref{app:examples} for more details). 
Suppose that $E=1.2$ and $\xi=2/3$. First, we numerically find that the optimal solution has $p^*_n = 0$ for all $n$, except for $n\in \{0,1,2\}$.  Then by invoking the KKT conditions, we find that  
\begin{align}
    p^*_0 & = \frac{\xi(5\xi+3E(1-\xi)-2)-1}{6\xi^2}, \\
p^*_1 & = \frac{1+\xi(2-3E+\xi)}{3\xi^2}, \\
p^*_2 & = \frac{(1+\xi)(\xi(3E-1)-1)}{6\xi^2},
\end{align}
with the following optimal value of the objective function $f(p)$ in \eqref{eq:quad_prog_trunc}:
\begin{align}
    f^* & = f(p^*) \\
    & = \frac{5+5\xi(2+\xi)-3E\xi(2+(2-E)\xi)}{6(1+\xi)^3}.
\end{align}

Using the KKT conditions, we find that the same solution is optimal for the quadratic program over an infinite number of variables in \eqref{eq:quad_prog_supp} (please see Appendix~\ref{app:examples} for more details). Moreover, for this example, the optimal state to distinguish ideal CV teleportation from its experimental implementation is given by 
\begin{align}\label{eq:opt-state-ex}
    \ket{\psi}_{RA} = \sqrt{p^*_0}\ket{00}_{RA}  + \sqrt{p^*_1}\ket{11}_{RA}  + \sqrt{p^*_2} \ket{22}_{RA}.
\end{align}


In summary, the energy-constrained channel fidelity between ideal teleportation and its experimental implementation can be calculated by employing the following three steps.
\begin{enumerate}
    \item
    Set the truncation parameter value $M$ to be larger than $E$. Then the quadratic program in \eqref{eq:quad_prog_trunc} can be solved numerically, which provides information about the non-zero elements in the optimal probability vector $p^*$.
    \item Use information about $p^*$ obtained in Step 1 to analytically solve the KKT conditions. If all the KKT conditions are satisfied, the solution obtained in Step 1 is also a solution to the quadratic program in \eqref{eq:quad_prog_supp}. If all the KKT conditions are not satisfied, repeat Step 1 with a larger value of the truncation parameter $M$ and then again solve the KKT conditions. 
    \item Use the solutions from Step 2 in \eqref{eq:tele-fid1}--\eqref{eq:quad-prog-cv-telport} to obtain analytical expressions for the energy-constrained channel fidelity between ideal CV teleportation and its experimental approximation and for the corresponding optimal state.
\end{enumerate}

\subsection{Experimental scheme for estimating the energy-constrained channel fidelity}

We now outline an experimental scheme
to estimate the energy-constrained channel fidelity between ideal teleportation and its experimental implementation. We note that the experimental procedure is particularly important when the value of the energy-constraint parameter is high.
\begin{enumerate}
    \item Alice experimentally prepares the state $\ket{\psi}_{RA}$,
    as defined in \eqref{eq:tps} with tunable parameters $\{p_n\}_n$. Throughout the experiment, we assume that $\sum_n p_n n \leq E$, where $E\in [0, \infty)$ is fixed. 
    \item Alice and Bob then perform the CV teleportation protocol \cite{BK98}. Depending on the
    squeezing in the shared TMSV state and the detection efficiency in the teleportation protocol, the final noisy state at Bob's end is given by $\rho_{RB} = (\I_R \otimes \T^{\xi}_{A \to B})(\psi_{RA})$.
    \item Bob estimates the fidelity between $\ket{\psi}_{RA}$ and $\rho_{RB}$ by using the quantum optical SWAP test  \cite{wang2001continuous,Filip02}.
    \item Depending on the fidelity value, Alice updates the parameters $\{p_n\}_n$ of the state $\ket{\psi}_{RA}$ and the CV teleportation experiment is repeated. Here, the goal is to update parameters such that the fidelity between $\psi_{RA}$ and $\rho_{RB}$ decreases. Since the objective function is convex (see Eq.~\eqref{eq:hessian-positivity}), an optimal solution can be obtained after a few iterations of the teleportation protocol.
\end{enumerate}  


\subsection{Comparison with previous results}

Let us compare our results with previous proposals based on the teleportation of coherent states and the two-mode squeezed vacuum state.
Let $\ket{\alpha}$ denote a coherent state and let $E = \vert \alpha \vert^2$. Then the fidelity between $\ket{\alpha}$ and $\mathcal{T}^{\xi}_{A \to B}(| \alpha\rangle\!\langle \alpha |)$ is given by
\begin{equation}
    F(| \alpha\rangle\!\langle \alpha |, \mathcal{T}^{\xi}_{A \to B}(| \alpha\rangle\!\langle \alpha |)) = \frac{1}{1+\xi}.
\end{equation}
Let $\ket{\psi_{\operatorname{TMS}}(E)}$ denote a two-mode squeezed vacuum state, defined as in \eqref{eq:tps} with $p_n = \frac{E^{n}}{(E+1)^{n+1}}$. Then we find that
\begin{equation}
    F(\psi_{\operatorname{TMS}}(E), \T^{\xi}_{A \to B}(\psi_{\operatorname{TMS}}(E))) = \frac{1}{1+ (2E+1)\xi}.
\end{equation}
These expressions can be evaluated using Eq.~(4.51) of \cite{AS17}. See Appendix~\ref{app:calcs-exp-states} for further details of these calculations.

\begin{figure}[ptb]
		\includegraphics[
		width=0.9\columnwidth
		]{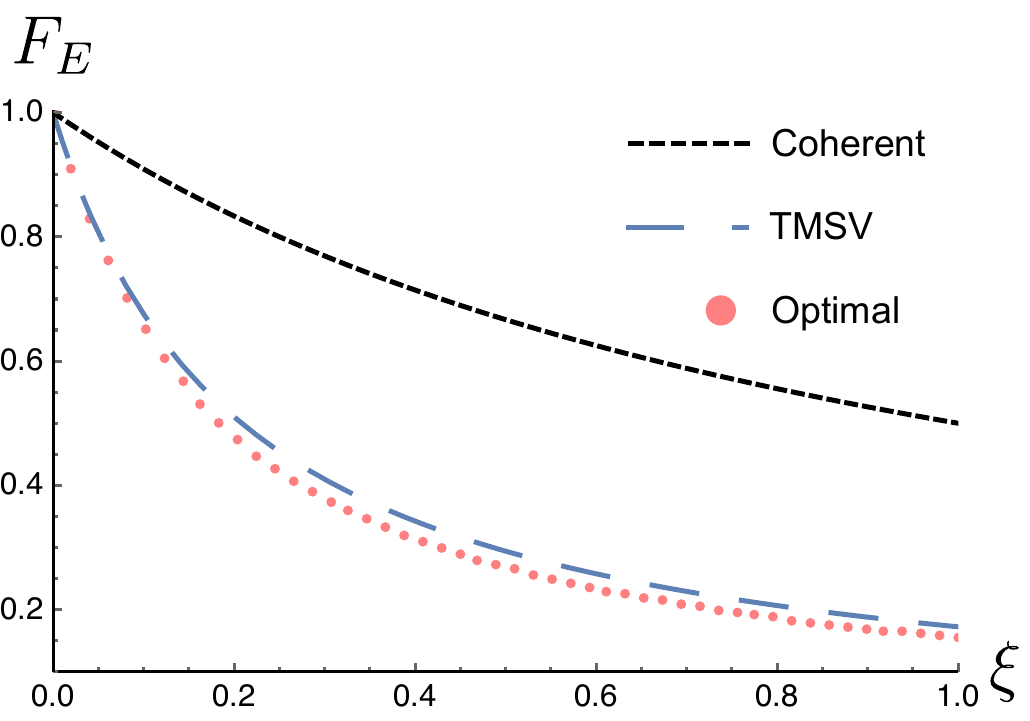}
	\caption{Plot of the fidelity between the identity channel $\mathcal{I}_{A \to B}$ and the additive-noise channel $\T^{\xi}_{A \to B}$ versus the additive noise parameter $\xi$, for different input states, with the choices indicated in the figure legend. We set the energy-constraint parameter value to $E=1.9$. Red dots plot the energy-constrained channel fidelity between $\I_{A \to B}$ and $\T^{\xi}_{A \to B}$. The black-dashed curve and the blue-dashed curve represent the fidelity of teleportation for a coherent state and a two-mode squeezed vacuum state, respectively.}
	\label{fig:comparison}
\end{figure}

In Figure~\ref{fig:comparison}, we plot the fidelity between $\I_{A\to B}$ and $\T^{\xi}_{A\to B}$ for several input states versus $\xi$, for $E=1.9$. We first numerically estimate the energy-constrained channel fidelity between $\I_{A \to B}$ and $\T^{\xi}_{A \to B}$ for $E=1.9$, which is plotted using red dots in Figure~\ref{fig:comparison}. In order to estimate the channel fidelity on a truncated Hilbert space, we set the truncation parameter $M = 50$. Black- and blue-dashed plots correspond to $F(| \alpha\rangle\!\langle \alpha |, \mathcal{T}^{\xi}_{A \to B}(| \alpha\rangle\!\langle \alpha |))$ and $F(\psi_{\operatorname{TMS}}(E), \T^{\xi}_{A \to B}(\psi_{\operatorname{TMS}}(E)))$, respectively.

Figure~\ref{fig:comparison} indicates that for a fixed value of $E$, coherent states and two-mode squeezed vacuum states are not optimal tests for the performance of CV quantum teleportation. Interestingly, however, the TMSV state is pretty close to being an optimal test for CV teleportation. From the prior results of Section~12 of \cite{SWAT}, it is known that the TMSV state is in fact optimal among all Gaussian input states, and so it is interesting that a non-Gaussian state achieves better performance.

\section{Quantifying the performance of experimental photodetectors}

In this section, we characterize the performance of an experimental approximation of an ideal photodetector.

Let $\P$ denote the channel corresponding to the ideal photodetector, whose action  on an input state $\rho$ is defined as follows: 
\begin{align}
\P(\rho) \equiv \sum_{n=0}^{\infty} \langle n \vert \rho \vert n \rangle | n\rangle\!\langle n |~.    
\end{align}
The interpretation of this channel is that it measures the input state in the photon-number basis and then outputs the measured value in a classical register. 

A simple way to model the noise in photodetection is to account for the loss of photons \cite{Loudon00}. In particular, we define the experimental approximation of $\P$ as follows:
\begin{align}
\widetilde{\P}^{\eta}(\rho) \equiv (\P \circ \L^{\eta}) (\rho)~.
\end{align}

We now evaluate the energy-constrained diamond distance between $\widetilde{\P}^{\eta}$  and $\P$. Employing the joint phase covariance of $\P$ and $\widetilde{\P}^{\eta}$, it follows that among all pure states, entangled superpositions of twin Fock states, as defined in \eqref{eq:tps}, are optimal to distinguish $\P$ from $\widetilde{\P}$ with respect to the energy-constrained diamond distance (see Appendix~\ref{app:photo-detector} for more details). 

Let $\{E \} \equiv E - \lfloor E \rfloor$. Then from  the direct-sum property of the trace distance on classical-quantum states and convexity of the function $x \to \eta^x$, where $\eta \in [0,1)$, we find that (see Appendix~\ref{app:photo-detector} for more details)
\begin{equation}
\frac{1}{2} \Vert \P - \widetilde{\P}^{\eta} \Vert_{\diamond E} = 1 -  \left[(1- \{E\}) \eta^{\lfloor E\rfloor} + \{E\}\eta^{\lceil E\rceil} \right]~.\label{eq:ecdd-pd-app-pd}
\end{equation}  
Moreover, the state that optimizes the energy-constrained diamond distance in \eqref{eq:ecdd-pd-app-pd} is given by the following mixed number state:
\begin{equation}\label{eq:opt-state}
\psi_A = (1- \{E\}) | \lfloor E \rfloor\rangle\!\langle \lfloor E \rfloor |_A + \{E\} | \lceil E \rceil\rangle\!\langle \lceil E \rceil |_A.
\end{equation}
From \eqref{eq:opt-state} it follows that entanglement is not necessary to attain the optimal distinguishability of the ideal photodetector $\P$ from its experimental approximation $\widetilde{\P}^{\eta}$. 
 
In Appendix~\ref{app:photo-detector}, we also consider a task of distinguishing two noisy photodetectors $\widetilde{\P}^{\eta_1}$ and $\widetilde{\P}^{\eta_2}$, and analytically calculate the energy-constrained sine distance between them. We find that it is given by
\begin{equation}
C_E(\widetilde{\P}^{\eta_1}, \widetilde{\P}^{\eta_2}) = \sqrt{1 - \left[(1- \{E\}) \mu^{\lfloor E\rfloor} + \{E\}\mu^{\lceil E\rceil} \right]^2} ~, 
\end{equation}
where
\begin{equation}
    \mu \equiv \sqrt{\eta_1 \eta_2} + \sqrt{(1-\eta_1)(1-\eta_2)}.
\end{equation}

\section{Conclusion}

In this paper, we proposed an optimal test to characterize the performance of CV teleportation in terms of the energy-constrained channel fidelity between ideal CV teleportation and its experimental approximation. 
Prior to our work, the accuracy in implementing CV teleportation was quantified by considering several input states, such as ensembles of coherent states, squeezed states, cat states, etc. 
We showed that, instead, entangled superpositions of twin Fock states are optimal to characterize the performance of CV teleportation. Thus, our result provides a benchmark for teleporting an unknown, energy-constrained state using CV quantum teleportation. Another interesting metric to quantify the accuracy in simulating ideal CV teleportation is the energy-constrained diamond distance between ideal CV teleportation and its experimental approximation. We leave the calculation of this quantity for future work, although we note here that it can be estimated by means of a semi-definite program \cite{W17} and truncation.

As an additional result, we analytically calculated the energy-constrained diamond distance between the ideal photodetector and its experimental approximation. Our main result here is that entanglement with a reference system is not required for quantifying the accuracy in implementing a photodetector and the optimal input state is a mixture of photon number states.  

There are a number of open questions to consider addressing in future work.
The Braunstein--Kimble (BK) protocol for CV teleportation \cite{BK98} can be generalized by using a non-Gaussian entangled state shared between Alice and Bob \cite{ban2002continuous,Dellanno2007continuous,Dellanno2010,Dellanno2013,Wang2015continuous,seshadreesan2015non,bose2018quantum}. It is an interesting open question to answer whether experimental realizations of CV teleportation with non-Gaussian resources can provide better teleportation fidelity than the BK protocol. Another interesting direction is to find applications of the techniques developed in Section III of \cite{sharma2020characterizing} to characterize the performance of these alternative strategies for CV quantum teleportation. We leave this for future work. Finally, it is an interesting open question to determine the optimal value of various energy-constrained distinguishability measures when the noise in a photodetector is modeled as a thermal noise channel.

\textit{Note on related work}: Recently, we became aware of related work of Lami \cite{Ludovico2020a}, in which he has established an upper bound on the energy-constrained diamond distance of the identity channel and an additive-noise channel by employing the methods of \cite{becker2020energy}.

In Ref.~\cite{SSW20}, we  have provided all source files (Mathematica and MATLAB) needed to generate the plots given in our paper. 

 \begin{acknowledgments}
This paper is dedicated to the memory of Jonathan~P.~Dowling. His enthusiasm for physics and unconditional support to students has always inspired us. May he never be forgotten.
We thank Ludovico Lami for helpful discussions, particularly for suggesting a simple proof for \eqref{eq:hessian-positivity}. KS and MMW acknowledge support from the National Science Foundation under Grant No.~2014010. BCS acknowledges support from NSERC and the Alberta Government.
 \end{acknowledgments}

\bibliographystyle{unsrt}
\bibliography{Ref}

\begin{thebibliography}{10}
\bibitem{Mathematica}
Mathematica and {MATLAB} files are available in the source files of our arxiv
  post.
\end{thebibliography}

\appendix

\section{Solving and evaluating the quadratic program for the fidelity of CV teleportation}

\label{sec:CV-teleportation_supp}

In this appendix, we show how to solve the quadratic program in \eqref{eq:quad_prog_supp}, by invoking the KKT conditions, and then we show details of the examples discussed in the main text. 



\subsection{Solving the quadratic program}

\label{app:solve-quad-prog}

For a finite number of variables, solutions to a quadratic program with inequality constraints can be obtained using a MATLAB package for quadratic programming, which employs the interior-point method \cite{pardalos1996interior}. Although the optimization problem in \eqref{eq:quad_prog_supp} is over an infinite number of variables, we argue below that for finite~$E$ and high values of the truncation parameter, it is sufficient to find solutions on a truncated Hilbert space. 

Let $M$ denote the truncation parameter, and let $\mathcal{H}_M$ denote an $(M+1)$-dimensional Fock space. 
Let $\varphi_A \in \mathcal{D}(\mathcal{H}_M)$, and let $\hat{\underline{n}}$ denote the truncated number operator:
\begin{equation}
 \hat{\underline{n}} = \sum_{n=0}^M n | n\rangle\!\langle n |. 
\end{equation}
Then from the mean energy constraint condition, we get $\tr(\hat{\underline{n}} \varphi_A) \leq E$. We define the energy-constrained channel fidelity between two quantum channels $\mathcal{N}_{A\to B}$ and $\mathcal{M}_{A\to B}$ on a truncated Hilbert space as 
\begin{multline}\label{eq:fid-truncated-supp}
    F_{E,M}(\mathcal{N}_{A\to B}, \mathcal{M}_{A\to B}) \\
   \equiv \inf_{\varphi_{RA} \in \mathcal{D}(\mathcal{H}_M^{\otimes 2}):\tr(\underline{\hat{n}}\varphi_A)\leq E} F(\mathcal{N}(\varphi_{RA}), \mathcal{M}(\varphi_{RA})),
\end{multline}
where $\varphi_{RA}$ is a purification of $\varphi_A$. Moreover, it is implicit that the identity channel acts on the reference system $R$. We further note that the following identity holds: 
\begin{multline}\label{eq:fid2-truncated-supp}
        F_{E,M}(\mathcal{N}_{A\to B}, \mathcal{M}_{A\to B})  \\
        \equiv \inf_{\varphi_{RA} \in \mathcal{D}(\mathcal{H}_M^{\otimes 2}):\tr(\hat{n}\varphi_A)\leq E} F(\mathcal{N}(\varphi_{RA}), \mathcal{M}(\varphi_{RA})),
\end{multline}
where we have replaced $\underline{\hat{n}}$ with $\hat{n}$, which follows from the fact that the reduced state of $\varphi_{RA} \in \mathcal{D}(\mathcal{H}^{\otimes 2}_M)$ on $A$ has support only on the truncated space. Moreover, \eqref{eq:fid2-truncated-supp} leads to the following identity: 
\begin{align}\label{eq:fid-upper-bound-supp}
 F_E( \mathcal{N}_{A\to B}, \mathcal{M}_{A\to B}) \leq F_{E,M}(\mathcal{N}_{A\to B}, \mathcal{M}_{A\to B}), 
\end{align}
which is a consequence of the optimization over a truncated space in \eqref{eq:fid2-truncated-supp} instead of an infinite-dimensional separable Hilbert space.

We now establish a lower bound on $F_E(\mathcal{N}_{A\to B}, \mathcal{M}_{A \to B})$ in terms of $F_{E,M}(\mathcal{N}_{A\to B}, \mathcal{M}_{A \to B})$, which, when combined with \eqref{eq:fid-upper-bound-supp}, will imply that solutions to \eqref{eq:quad_prog_supp} can be obtained by solving a quadratic program on a truncated Hilbert space. 
For completeness we first argue that the set of density operators acting on a truncated Hilbert space with a finite mean energy constraint is dense in the set of density operators acting on an infinite-dimensional Hilbert space and with the same mean energy constraint \cite{sharma2020characterizing}.  Let $\rho_{RA}$ denote a density operator acting on an infinite-dimensional separable Hilbert space, such that $\tr(\hat{n}_A \rho_{RA}) \leq E$, where $E>0$. Let $\Pi_A^M$ denote an $(M+1)$-dimensional projector defined as
\begin{align}
\Pi^M_A = \sum_{n=0}^{M} | n\rangle\!\langle n |~. 
\end{align}	
Then from \cite[Section III E]{sharma2020characterizing}, it follows that 
\begin{align}\label{eq:expec-trunc-projector}
    \tr(\Pi^M_A \rho_{RA}) \geq 1 - \frac{E}{M+1}.
\end{align}
Let $\rho^M_{RA}$ denote the following truncated state:
\begin{align}
    \rho^{M}_{RA} = \frac{\Pi^M_A \rho_{RA} \Pi^M_A}{\tr(\Pi^M_A \rho_{RA})}.
\end{align}
Then by invoking the gentle measurement lemma \cite{W99, ON07}, we get 
\begin{align}\label{eq:fid-trunc}
    F(\rho_{RA}, \rho^M_{RA}) \geq 1 - \frac{E}{M+1},
\end{align}
which implies that the fidelity between the truncated state $\rho^M_{RA}$ and $\rho_{RA}$ is close to one for low values of $E$ and high values of the truncation parameter $M$. For the version of the gentle measurement lemma that we employ, please see the proof of Lemma~9.4.1 of \cite{W15book}.

We define the energy-constrained sine distance between two channels $\N_{A\to B}$ and $\mathcal{M}_{A \to B}$ on a truncated Hilbert space as follows: 
\begin{multline}
    C_{E, M}(\mathcal{N}_{A\to B}, \mathcal{M}_{A\to B}) 
   \equiv \\ \sup_{\varphi_{RA} \in \mathcal{D}(\mathcal{H}_M^{\otimes 2}):\tr(\hat{n}\varphi_A)\leq E} \sqrt{1-F(\mathcal{N}(\varphi_{RA}), \mathcal{M}(\varphi_{RA}))}~.
\end{multline}

Then from \eqref{eq:expec-trunc-projector}, \eqref{eq:fid-trunc}, and arguments similar to those used in \cite[Proposition~2]{sharma2020characterizing}, we establish the following inequalities:
\begin{align}
  & C_{E,M}(\mathcal{N}_{A\to B}, \mathcal{M}_{A\to B})\notag \\
  & \leq  C_E(\mathcal{N}_{A\to B}, \mathcal{M}_{A\to B})
  \\ & \leq 2 \sqrt{\frac{E}{M+1}} + C_{E,M}(\mathcal{N}_{A\to B}, \mathcal{M}_{A\to B}).
\end{align}

Finally, by squaring the inequality on the right side and from a simple rearrangement, we get 
\begin{multline}\label{eq:fid-inequalities-supp}
    F_E(\mathcal{N}_{A\to B}, \mathcal{M}_{A\to B}) 
    \\
    \geq 1 - \left(2\sqrt{\frac{E}{M+1}} + \sqrt{1 - F_{E,M}(\mathcal{N}_{A\to B}, \mathcal{M}_{A\to B})} \right)^2,
\end{multline}
which leads to the desired result by combining with \eqref{eq:fid-upper-bound-supp}
\begin{multline}
  1 - \left(2\sqrt{\frac{E}{M+1}} + \sqrt{1 - F_{E,M}(\mathcal{N}_{A\to B}, \mathcal{M}_{A\to B})} \right)^2  \\
  \leq   F_E( \mathcal{N}_{A\to B}, \mathcal{M}_{A\to B}) \leq F_{E,M}(\mathcal{N}_{A\to B}, \mathcal{M}_{A\to B}), 
\end{multline} 
 In other words, for low values of the mean energy constraint $E$, the energy-constrained channel fidelity between two quantum channels $\mathcal{N}$ and $\mathcal{M}$ can be estimated with arbitrary accuracy by using the energy-constrained channel fidelity on a truncated input Hilbert space with sufficiently high values of the truncation parameter $M$.
Moreover, for low values of $E$ and high values of $M$, solutions to the quadratic program in \eqref{eq:quad_prog_supp} can be obtained by solving the following quadratic program:
\begin{align}\label{eq:quad_prog_truncated_supp}
 &F_{E,M}(\I_A, \T^{\xi}_A)\nonumber \\
 &= \left\{ \begin{array}{l l}
    \inf_{p} &  f(p) \\[6pt]
\text{subject to} & \sum_{n=0}^{M} n p_n \leq E\\[3pt]
& -p_n \leq 0, \forall n \in \{0, 1, \dots, M-1, M\}  \\[3pt]
& \sum_{n=0}^{M} p_n = 1,
\end{array}\right.
\end{align}
where $p=(p_0, p_1, \dots, p_M)$ and $F_{E,M}(\mathcal{I}_A, \mathcal{T}^{\xi}_A)$ is given by \eqref{eq:fid2-truncated-supp}. 
It is easy to check that \eqref{eq:quad_prog_truncated_supp} is equivalent to the primal optimization problem in \eqref{eq:kkt-primal}.

We now elaborate the argument of \cite{Ludovico2020} that the function $f(p)$ is convex in $p$ (an abbreviated version of this argument was already presented in the main text). The Hessian matrix corresponding to the objective function $f(p)$ is given by 
\begin{align}
    & A(\xi) \notag \\
    & =2\sum_{n,m=0}^M  \sum_{k=0}^{\min\{n,m\}} \binom{n}{k}\binom{m}{k} \frac{\xi^{2k}}{1+\xi}\left(\frac{1}{1+\xi}\right)^{n+m}| n\rangle\!\langle m | \label{eq:hessian-1}\\
    &= 2 \sum_{k=0}^M \frac{\xi^{2k}}{1+\xi} \left[\sum_{n=0}^M \binom{n}{k} \frac{1}{(1+\xi)^n}\ket{n}\right]\times \notag \\
    & \qquad\qquad\left[\sum_{m=0}^M \binom{m}{k} \frac{1}{(1+\xi)^m}\bra{m} \right]\\
    & = 2  \sum_{k=0}^M \frac{\xi^{2k}}{1+\xi} | \Upsilon\rangle\!\langle\Upsilon |,
\end{align}
where 
\begin{align}
    \ket{\Upsilon} = \sum_{n=0}^M \binom{n}{k} \frac{1}{(1+\xi)^n}\ket{n}~.
    \label{eq:hessian-4}
\end{align}
Thus we get 
\begin{align}
\langle \Phi \vert A(\xi) \vert \Phi\rangle = 2 \sum_{k=0}^M \frac{\xi^{2k}}{1+\xi} \vert \langle \Phi \vert \Upsilon \rangle \vert^2 \geq 0~, 
\end{align}
which implies that the objective function $f(p)$ in \eqref{eq:quad_prog_truncated_supp} is convex in $p$. Since the aforementioned proof holds for any value of the truncation parameter $M$, it further implies that the objective function in \eqref{eq:quad_prog_supp} is also convex.

Finally, note that the inequality constraints in \eqref{eq:quad_prog_supp} are linear, which implies that the Lagrangian
\begin{multline}
    L(p, \mu, \beta, \gamma) = f(p) + \mu \left(\sum_n n p_n - E\right) \\
    - \sum_n \beta_n p_n + \gamma \left(\sum_n p_n -1\right) 
\end{multline}
is also convex in $p$, where we introduced the dual variables $\mu, \gamma$, and $\beta_n$, for $n\in \mathbb{Z}^{\geq0}$, similar to those in \eqref{eq:lag}. 

\subsection{Solutions of the quadratic program for some examples}

\label{app:examples}

We now provide solutions to \eqref{eq:quad_prog_supp} for several examples.

\subsubsection{First example: $E=0.6$ and $\xi=0.25$}

Suppose that $E=0.6$ and $\xi=0.25$. We first numerically find the optimal solution to \eqref{eq:quad_prog_truncated_supp} using the MATLAB package for quadratic programming \cite{Mathematica}.
Furthermore, we find numerically that for this case, the optimal value $f^*$ of the objective function and the corresponding optimal solution $p^*$ do not change for values of the truncation parameter from $M=1$ to $M=50$. We provide a reasoning for this result by analytically solving the quadratic program by invoking the KKT conditions. We note that if the optimal value of the objective function does not change with an increment in the truncation parameter, it implies that the same solution should be optimal for the infinite-dimensional optimization problem in \eqref{eq:quad_prog_supp}. 

We now analytically find primal and dual feasible points that satisfy the KKT conditions and hence the optimal solution to \eqref{eq:quad_prog_supp} for $E=0.6$ and $\xi=0.25$.
We begin by presenting the KKT conditions for the optimization problem in \eqref{eq:quad_prog_supp}. 
\begin{equation}
    \begin{array}{ll} \text{Stationarity}
         &  \partial_p L(p, \mu, \beta, \gamma)\vert_{p=p^*} = 0  \\[6pt]
       \text{Primal feasibility} & \sum_n n p^*_n - E \leq 0  \\[3pt]
       & p^*_n \geq 0, \forall n \in \mathbb{Z}^{\geq 0}\\[3pt]
       & \sum_{n} p^*_n = 1\\[6pt]
        \text{Dual feasibility} & \mu \geq 0  \\[3pt]
         & \beta_n \geq 0, \forall n \in \mathbb{Z}^{\geq 0}\\[6pt]
\text{Complementary slackness} & \mu \left(\sum_n n p^*_n - E\right) = 0 \\[3pt]
& \beta_n p^*_n = 0, \forall n \in  \mathbb{Z}^{\geq 0}
    \end{array}
\end{equation}\label{eq:kkt-cond}

 We find a set $(\tilde{p}, \tilde{\mu},\tilde{\beta},\tilde{\gamma})$ that satisfies the KKT conditions, which further implies that $\tilde{p} = p^*$ and $f^* = f(p^*)$, where $f(p)$ is defined in \eqref{eq:quad_prog_supp}. Let $\tilde{p}_n=0$ for all $n \geq 2$. Let us assume that the mean energy constraint is saturated, i.e., $\tilde{p}_1=E$, which satisfies one of the complementary slackness conditions. To satisfy $\beta_n \tilde{p}_n=0$, for $n=0,1$, we set $\beta_0=\beta_1=0$. From the primal feasibility condition, we get $\tilde{p}_0 = 1-\tilde{p}_1 = 1 -E$. All we need to show now is that $\mu\geq 0$ and $\beta_n\geq 0$ for all $n\geq2$. Combining all these facts will imply that the KKT conditions are satisfied. First, to simplify the calculation, we define 
\begin{align}\label{eq:gnm}
G(n, m, \xi)    = \sum_{k=0}^{\min\{n,m\}} \binom{n}{k}\binom{m}{k} \frac{\xi^{2k}}{(1+\xi)^{n+m+1}}~. 
\end{align}

Then from the stationarity condition $\tilde{p}_0$ and $\tilde{p}_1$, we get the following linear system of equations: 
\begin{align}
    2 G(0,0, \xi)\tilde{p}_0 + 2 G(1, 0, \xi) \tilde{p}_1 + \gamma &=0,\\
    2 G(0,1,\xi)\tilde{p}_0 + 2 G(1,1,\xi)\tilde{p}_1 + \mu + \gamma & =0.
\end{align}
By solving for $\mu$ and $\gamma$, we find \cite{Mathematica}
\begin{align}
    \mu &= \frac{2 \xi(1 - (2E-1)\xi)}{(1+\xi)^3} 
    > 0~,\\
    \gamma & = -\frac{2(1+(1-E)\xi)}{(1+\xi)^2},
\end{align}
for $E=0.6$ and $\xi = 0.25$. 

Since $\mu>0$, in order to satisfy all the KKT conditions for \eqref{eq:quad_prog_supp}, we only need to show that $\beta_n\geq 0$ for all $n \geq 2$.
From the stationarity condition for $\tilde{p}_n$ we get, 
\begin{align}\label{eq:betan}
  \beta_n =  2 G(0,n,\xi)\tilde{p}_0 + 2G(1,n,\xi)\tilde{p}_1 + n\mu + \gamma ~.
\end{align}
The only negative term in \eqref{eq:betan} is $\gamma$. For this example we get $\mu= 0.2432$ and $\gamma=-1.408$. Since $-\gamma/\mu =5.79$, we get $n \mu \geq -\gamma$, $\forall n\geq 6$. This further implies that $\beta_n \geq 0$, $\forall n\geq 6$. Moreover, we solve for $\beta_2$, $\beta_3$, $\beta_4$, and $\beta_5$ using \eqref{eq:betan} and find their values of be $0.041$, $0.1160$,  $0.2202$, and $0.348$, respectively \cite{Mathematica}. Thus we get $\beta_n \geq 0$  for all $n\geq 2$. This completes the proof. 

Since all the KKT conditions are satisfied and since $f(p)$ is a convex function, we conclude that $\tilde{p}$ is the optimal solution, i.e., $\tilde{p}= p^* = (1-E, E, 0, \dots, 0)$ and the optimal objective function value is given by 
\begin{align}
    f^* & = f(p^*) \\
    & = \frac{1+ \xi(2+\xi-2E(1+(1-E)\xi))}{(1+\xi)^3} \\
    & = 0.6310,
\end{align}
for $E=0.6$ and $\xi = 0.25$, which is equal to the optimal value obtained numerically. Furthermore, for this case, the optimal state corresponding to the channel fidelity between the ideal teleportation and its experimental approximation is given by 
\begin{align}
    \ket{\psi}_{RA} = \sqrt{1-E}\ket{0}_A\ket{0}_R + \sqrt{E}\ket{1}_A \ket{1}_R~. 
\end{align}

\subsubsection{Second example: $E=1.2$ and $\xi=2/3$}

Let us consider the case when $E=1.2$ and $\xi=2/3$. Numerically, we find that the optimal solution has $\tilde{p}_n=0$ for all $n$, except for $n \in \{0,1,2\}$. To satisfy the complementary slackness condition, we set $\beta_0 =  \beta_1 = \beta_2 = 0$. Similar to the previous case, we assume that the energy constraint is satisfied, i.e., 
    $\tilde{p}_1 + 2 \tilde{p}_2 = E.$
    By invoking the stationarity conditions for $\tilde{p}_0$, $\tilde{p}_1$, and $\tilde{p}_2$, and the primal feasibility condition and by solving the linear system of equations, we get that \cite{Mathematica}
\begin{align}
\tilde{p}_0 &= \frac{\xi(5\xi+3E(1-\xi)-2)-1}{6\xi^2} > 0,\\
\tilde{p}_1 & = \frac{1+\xi(2-3E+\xi)}{3\xi^2} \geq 0,\\
\tilde{p}_2 & = \frac{(1+\xi)(\xi(3E-1)-1)}{6\xi^2} > 0,\\
\mu & = \frac{\xi(1+(1-E)\xi)}{(1+\xi)^3}> 0,\\
\gamma &= -\frac{5+(5-3E)\xi}{3(1+\xi)^2},
\end{align}
for $E=1.2$ and $\xi=2/3$. Moreover, similar to the previous case, we find that $\beta_n \geq 0$  for all $n \geq 3$ \cite{Mathematica}. 

Since all the KKT conditions are satisfied and since $f(p)$ is a convex function, we conclude that $\tilde{p}$ is the optimal solution, i.e., $\tilde{p} = p^*$. Moreover, the optimal objective function is given by 
\begin{align}
    f^* = f(p^*) = \frac{5+5\xi(2+\xi)-3E\xi(2+(2-E)\xi)}{6(1+\xi)^3}~,
\end{align}
and the corresponding optimal state to distinguish the ideal teleportation channel from its experimental approximation is
\begin{align}
    \ket{\psi}_{RA} = \sqrt{\tilde{p}}_0\ket{0}_A \ket{0}_R + \sqrt{\tilde{p}}_1\ket{1}_A \ket{1}_R + \sqrt{\tilde{p}}_2\ket{2}_A \ket{2}_R.
\end{align}

\subsubsection{Third example: $\xi$ close to zero}

We provide an analytical solution for another interesting example when $\xi$ is close to zero, which corresponds to the case of the additive-noise channel converging to the ideal teleportation channel. This example is experimentally relevant, as the goal of an approximate teleportation protocol is to converge to the ideal teleportation channel. In such a scenario we argue that 
\begin{multline}\label{eq:Ranjith-state}
    \ket{\psi}_{RA} = \sqrt{1- \{E\}} \ket{\lfloor E\rfloor}_A \ket{\lfloor E\rfloor}_R \\
    + \sqrt{\{E\}} \ket{\lceil E \rceil}_A \ket{\lceil E \rceil}_R
\end{multline}
is the optimal state for the optimization problem in \eqref{eq:tele-fid1}, where $\{E\} = E - \lfloor E \rfloor$. Moreover, the minimum fidelity in \eqref{eq:tele-fid1} is given by 
\begin{multline}
      F_E(\I_A, \T^{\xi}_A) 
      =  (1- \{E\})^2 G(\lfloor E \rfloor, \lfloor E \rfloor, \xi)\\
      +2\{E\}(1-\{E\}) G(\lfloor E \rfloor, \lceil E \rceil, \xi) \\
      + (\{E\})^2  G(\lceil E \rceil,\lceil E \rceil,\xi), 
\end{multline}
where $G(n, m, \xi)$ is defined in \eqref{eq:gnm}.

Since the marginal of the state $\ket{\psi}_{RA}$ in \eqref{eq:Ranjith-state} has energy $\tr(\hat{n} \psi_A) = E$, one of the complementary slackness conditions is satisfied. We assume that $p^*_n=0$ for all $n$, except for $n\in \{\lfloor E \rfloor, \lceil E \rceil\}$.
Therefore, $p^*_{\lfloor E \rfloor} = 1 - \{E\}$ and $p^*_{\lceil E \rceil} = \{E\}$, which implies that $\beta_{\lfloor E\rfloor}= \beta_{\lceil E \rceil} = 0$, to satisfy other complementary slackness conditions.  Similar to the previous examples, we need to solve the following linear system of equations and show that $\mu \geq 0$ when $\xi$ is close to zero:  
\begin{widetext}
\begin{align}
    2 p^*_{\lfloor E\rfloor}  G(\lfloor E \rfloor, \lfloor E \rfloor, \xi) + 2 p^*_{\lceil E \rceil}  G(\lfloor E \rfloor, \lceil E \rceil, \xi) + \lfloor E \rfloor \mu + \gamma & = 0,\\
    2 p^*_{\lceil E\rceil}  G(\lceil E \rceil, \lceil E \rceil, \xi) + 2 p^*_{\lfloor E \rfloor} G(\lfloor E \rfloor, \lceil E \rceil, \xi)  + \lceil E \rceil \mu + \gamma &= 0 . 
\end{align}
By solving for $\mu$, we get 
\begin{align}
\mu &= 2\left( p^*_{\lfloor E \rfloor} (G(\lfloor E \rfloor, \lfloor E \rfloor,\xi) - G(\lfloor E \rfloor, \lceil E \rceil, \xi ))+ p^*_{\lceil E \rceil} ( G(\lfloor E \rfloor, \lceil E \rceil, \xi ) - G(\lceil E \rceil, \lceil E \rceil, \xi))   \right),\\
& \approx 2 \xi + O(\xi^2)~,
\end{align}
\end{widetext}
which implies that the leading order term $2\xi \geq 0$ 
for any finite value of the energy constraint $E$. Similarly, we find that 
\begin{align}
    \beta_n \approx (\lfloor E \rfloor - n) (\lceil E \rceil - n)\xi^2 + O(\xi^3), \forall n,
\end{align}
where again the leading term implies that $\beta_n\geq 0$ when $\xi$ is close to zero. By combining everything, we conclude that all the KKT conditions are satisfied. Hence, for $\xi$ close to zero, the state in \eqref{eq:Ranjith-state} is optimal for the task of distinguishing the ideal teleportation channel from its experimental approximation when there is a finite energy constraint on the input states to the channels. 

\subsection{Other calculations for experimentally relevant  states}

\label{app:calcs-exp-states}

Here we calculate the fidelity of teleportation for several experimentally relevant quantum states with energy constraints, such as  coherent states and the two-mode squeezed vacuum (TMSV) state. Let $\ket{\alpha}$ denote a coherent state and let $E = \vert \alpha \vert^2$. We note that the covariance matrix of a coherent state is a two-dimensional identity matrix, which under an additive-noise channel $\T^{\xi}_{A \to B}$ becomes $V_{\T^{\xi}_{A \to B}(| \alpha\rangle\!\langle \alpha |)} = \text{diag}(1+2\xi)$. Therefore, we get~\cite{AS17}
\begin{align}
    F(| \alpha\rangle\!\langle \alpha |,\T^{\xi}_{A \to B}(| \alpha\rangle\!\langle \alpha |)) & = \frac{2}{\sqrt{\text{Det}(\text{diag}(2(1+\xi)))}} \notag \\
    & = \frac{1}{1+\xi}. 
\end{align}

On the other hand, the covariance matrix of a two-mode squeezed vacuum state is given by 
\begin{align}
    V_{\psi_{\operatorname{TMS}}(\bar{n})} = \begin{bmatrix}
    (2 \bar{n}+1) I_2 & 2\sqrt{\bar{n}(\bar{n}+1)} \sigma_z\\
    2\sqrt{\bar{n}(\bar{n}+1)} \sigma_z & (2 \bar{n}+1) I_2
    \end{bmatrix},
\end{align}
which transforms as 
\begin{align}
        V_{\T^{\xi}(\psi_{\operatorname{TMS}}(\bar{n}))} = \begin{bmatrix}
    (2 \bar{n}+1) I_2 & 2\sqrt{\bar{n}(\bar{n}+1)} \sigma_z\\
    2\sqrt{\bar{n}(\bar{n}+1)}\sigma_z & (2 \bar{n}+2\xi+1) I_2
    \end{bmatrix}. 
\end{align}
Then the fidelity between $\psi_{\operatorname{TMS}}(\bar{n})$ and $\T^{\xi}(\psi_{\operatorname{TMS}}(\bar{n}))$ is given by \cite{AS17}
\begin{align}
&F(\psi_{\operatorname{TMS}}(\bar{n}), \T^{\xi}(\psi_{\operatorname{TMS}}(\bar{n})))\nonumber\\
&= \frac{4}{\sqrt{\text{Det}(V_{\psi_{\operatorname{TMS}}(\bar{n})}+ V_{\T^{\xi}(\psi_{\operatorname{TMS}}(\bar{n}))})}} \\
& = \frac{1}{1+ (2\bar{n}+1)\xi}~.
\end{align}

\section{Calculations for the approximation of a photodetector}\label{app:photo-detector}

In this appendix, we provide a proof for the analytical form of the energy-constrained diamond distance between the ideal photodetector and its experimental approximation. Let $\mathcal{P}$ and $\widetilde{\mathcal{P}}^{\eta}$ denote the ideal photodetector and the noisy photodetector, respectively. Moreover, $\mathcal{P}$ and $\widetilde{\mathcal{P}}^{\eta}$ transform an input state $\rho$ as follows: 
\begin{align}
   \P(\rho) &= \sum_{n=0}^{\infty} \langle n \vert \rho \vert n \rangle | n\rangle\!\langle n |~,\label{def:ideal-pd-supp}\\
   \widetilde{\P}^{\eta}(\rho) & =  (\P \circ \L^{\eta}) (\rho)~,
\end{align}
where $\L^{\eta}$ denotes a pure-loss channel with transmissivity~$\eta\in [0,1)$.

Let 
\begin{align}\label{eq:tps-pd}
\ket{\psi}_{RA}  = \sum_n \sqrt{p_n} \ket{n}_R \ket{n}_A,
\end{align}
where $\sum_n p_n =1$ and $\sum_n n p_n \leq E$. 

Consider the following chain of equalities:
\begin{widetext}
\begin{align}
&\left\Vert (\I_R\otimes \P_A)(\psi_{RA}) -(\I_R \otimes \widetilde{\P}^{\eta}_A)(\psi_{RA})\right\Vert_1 \nonumber \\
&= \left\Vert  (\I_R\otimes \P_A)(\psi_{RA}) -(\I_R \otimes (\P_A\circ \L^{\eta}_A))(\psi_{RA})\right\Vert_1\\
& = \left\Vert \sum_{n,m=0}^{\infty}  \sqrt{p_m p_n}| n\rangle\!\langle m |_R \otimes \left[\P(| n\rangle\!\langle m |_A)  -   \sum_{k=0}^{\text{min}\{n,m\}} \sqrt{\binom{n}{k}\binom{m}{k}\eta^{n+m - 2k}(1-\eta)^{2k}}  \P(\ket{n-k}\bra{m-k}_A)\right]  \right\Vert_1\\
& = \left\Vert  \sum_{n=0}^{\infty} p_n| n\rangle\!\langle n |_R \otimes \left[| n\rangle\!\langle n |_A  -   \sum_{k=0}^{n} \binom{n}{k} \eta^{k}(1-\eta)^{n-k} | k \rangle \! \langle k |_A \right]  \right\Vert_1\\
& = \sum_{n=0}^{\infty} p_n \left\Vert | n\rangle\!\langle n |_A -  \sum_{k=0}^{n} \binom{n}{k} \eta^{k}(1-\eta)^{n-k}  | k \rangle \! \langle k |_A \right\Vert_1\\
& = \sum_{n=0}^{\infty} p_n \left[\sum_{k=0}^{n-1}  \binom{n}{k} \eta^{k}(1-\eta)^{n-k}    +  1- \eta^{n}    \right]\\
& = 2\left(1-  \sum_{n=0}^{\infty} p_n  \eta^{n}\right)~.
\end{align}
\end{widetext}
The second equality follows from the action of a pure-loss channel on a number state. The third equality follows from \eqref{def:ideal-pd-supp}. The fourth equality is a consequence of the direct-sum property of the trace distance on classical-quantum states. The rest of the steps follow from basic algebraic manipulations.

Therefore, the energy-constrained diamond distance between $\P$ and $\widetilde{\P}^{\eta}$ reduces to the following optimization problem:
\begin{multline}\label{eq:dd-pd}
\frac{1}{2}\Vert \P - \widetilde{\P}^{\eta}\Vert_{\diamond E} =\\ 
\max_{\{p_n\geq 0\}_n: \sum_n p_n = 1, \sum_{n} n p_n \leq E } \left(1- \sum_{n=0}^{\infty} p_n \eta^n\right).
\end{multline}

The optimization in \eqref{eq:dd-pd} can be solved by following a method introduced in \cite{N18}. We provide a proof for completeness. Suppose that
\begin{align}
F & = \sum_n n p_n, \\
A_l & = \sum_{n=0}^{\lfloor F \rfloor} p_n, \\
A_u  & = \sum_{n=\lceil F \rceil}^{\infty}p_n, \\
F_l & = \sum_{n=0}^{\lfloor F\rfloor} (p_n/A_l) n,\\
F_u & = \sum_{n=\lceil E \rceil}^{\infty} (p_n/A_u) n.
\end{align}
Then it follows that 
\begin{align}
A_l&+A_u  = 1,\\
F &= A_l F_l + A_u F_u,\\
F_l &\leq \lfloor F \rfloor, \\
F_u & \geq \lceil F \rceil. 
\end{align}

Consider the following chain of inequalities: 
\begin{align}
\sum_{n=0}^{\infty} p_n \eta^n &= A_l \sum_{n=0}^{\lfloor F \rfloor} \frac{p_n}{A_l} \eta^n + A_u \sum_{n=\lceil F \rceil}^{\infty} \frac{p_n}{A_u} \eta^n\\
& \geq A_l \eta^{F_l} + A_u \eta^{F_u}\\
& \geq p_{\lfloor F \rfloor}\eta^{\lfloor F\rfloor} + p_{\lceil F \rceil} \eta^{\lceil F \rceil}\\
& = (1- \{F\})\eta^{\lfloor F\rfloor} + \{F\} \eta^{\lceil F \rceil}.\label{eq:opt-func}
\end{align}
where the first inequality follows from the convexity of the function $x \to \eta^x$. The last inequality follows from the fact that the chord joining $(\lfloor F \rfloor, \eta^{\lfloor F \rfloor})$ and $(\lceil F \rceil, \eta^{\lceil F \rceil})$ is below the chord joining $(A_l, \eta^{F_l})$ and $(A_u, \eta^{F_u})$ due to the convexity of the function. Moreover, the energy of the initial state can be satisfied by taking $F_l = \lfloor F \rfloor$ and $F_u = \lceil F \rceil$, and the corresponding probability elements are given by $ p_{\lfloor F \rfloor} = 1- \{F\}$ and $p_{\lceil F \rceil} = \{F\}$. 

Since \eqref{eq:opt-func} monotonically decreases with $F$, it implies that the solution to the optimization problem in \eqref{eq:dd-pd} is given by a state that saturates the energy constraint. Therefore, the optimal state is given by 
\begin{equation}\label{eq:opt-state-pd-supp}
\psi_A = (1- \{E\}) | \lfloor E \rfloor\rangle\!\langle \lfloor E \rfloor |_A + \{E\} | \lceil E \rceil\rangle\!\langle \lceil E \rceil |_A~,
\end{equation}
which leads to 
\begin{equation}
\frac{1}{2} \Vert \P - \widetilde{\P}^{\eta} \Vert_{\diamond E} = 1 -  \left[(1- \{E\}) \eta^{\lfloor E\rfloor} + \{E\}\eta^{\lceil E\rceil} \right]~.
\end{equation}  

We now provide proof details for the energy-constrained sine distance between two noisy photodetectors $\widetilde{\P}^{\eta_1}$ and $\widetilde{\P}^{\eta_2}$. 
First note that the output of a noisy photodetector $\widetilde{\P}^{\eta}$ when the input state is $\psi_{RA}$ in \eqref{eq:tps-pd} is given by
\begin{align}
\widetilde{\P}^{\eta}(\psi_{RA}) & = (\P\circ \L^{\eta})(\psi_{RA}) \\
& = \sum_{n}\sum_{k=0}^{n} \beta(p_n, \eta, k) | n\rangle\!\langle n |_R \otimes| k \rangle \! \langle k |_A,
\end{align}
where 
\begin{align}
    \beta(p_n, \eta, k) = p_n  \binom{n}{k} \eta_1^k (1-\eta_1)^{n-k}
\end{align} 
are the eigenvalues of $\widetilde{\P}^{\eta}(\psi_{RA})$. Therefore, the fidelity between $\widetilde{\P}^{\eta_1}(\psi_{RA})$ and $\widetilde{\P}^{\eta_2}(\psi_{RA})$ is given by
\begin{align}
& F(\widetilde{\P}^{\eta_1}(\psi_{RA}), \widetilde{\P}^{\eta_2}(\psi_{RA})) \notag \\
&= \left[\tr\left(\sqrt{\sqrt{\widetilde{\P}^{\eta_1}(\psi_{RA})} \widetilde{\P}^{\eta_2}(\psi_{RA}) \sqrt{\widetilde{\P}^{\eta_1}(\psi_{RA})} }\right)\right]^2\\
& = \left[\sum_{n}\sum_{k=0}^n \sqrt{ \beta(p_n, \eta_1, k)\beta(p_n, \eta_2, k)  } \right]^2\\
& = \left[\sum_n p_n \sum_{k=0}^n \binom{n}{k} \sqrt{\eta_1 \eta_2}^k \Big[(1-\eta_1)(1-\eta_2)\Big]^{n-k} \right]^2\\
& =\left[\sum_n p_n \Big(\sqrt{\eta_1 \eta_2} + \sqrt{(1-\eta_1)(1-\eta_2)} \Big)^n  \right]^2\\
& = \left[\sum_n p_n \mu^n \right]^2,
\end{align}
where
\begin{align}
    \mu = \sqrt{\eta_1 \eta_2} + \sqrt{(1-\eta_1)(1-\eta_2)}. 
\end{align}

Then from \cite{N18}, we find that the energy-constrained sine distance between $\widetilde{\P}^{\eta_1}$ and $\widetilde{\P}^{\eta_2}$ is given by
\begin{align}\label{eq:sined-pd2-supp}
C_E(\widetilde{\P}^{\eta_1}, \widetilde{\P}^{\eta_2}) = \sqrt{1 - \left[(1- \{E\}) \mu^{\lfloor E\rfloor} + \{E\}\mu^{\lceil E\rceil} \right]^2} ~, 
\end{align}
where $\{E \} = E - \lfloor E \rfloor$. Moreover, the state that optimizes the energy-constrained sine  distance in \eqref{eq:sined-pd2-supp} is given by \eqref{eq:opt-state-pd-supp}, which proves that entanglement is not necessary for the optimal distinguishability of two noisy photodetectors.

\end{document}